\documentclass[12pt,english]{article}
\usepackage[english]{babel}
\usepackage{amsmath,amssymb,amscd,color}
\usepackage{amsfonts}
\usepackage{mathrsfs}
\usepackage{mathtools}
\usepackage{graphicx}
\usepackage{url}
\usepackage{hyperref}
\usepackage{cite}
\usepackage{physics}
\usepackage{bbold}
\usepackage{subcaption}
\usepackage{braket}
\usepackage{verbatim}
\usepackage[font=footnotesize,labelfont=bf,width=.9\textwidth]{caption}
\usepackage{multirow}
\usepackage{boldline}
\usepackage{enumitem}
\usepackage{placeins}
\usepackage{ctable,tabularx}
\usepackage{longtable}
\allowdisplaybreaks

\hypersetup{
    colorlinks,%
    citecolor=blue,%
    filecolor=blue,%
    linkcolor=blue,%
    urlcolor=blue
}

\makeatletter
\renewcommand\section{\@startsection {section}{1}{\z@}%
                                   {-5.5ex \@plus -1ex \@minus -.2ex}
                                   {2.3ex \@plus.2ex}%
                                   {\normalfont\large\bfseries}}
\renewcommand\subsection{\@startsection{subsection}{2}{\z@}%
                                     {-3.25ex\@plus -1ex \@minus -.2ex}%
                                     {1.5ex \@plus .2ex}%
                                     {\normalfont\bfseries}}

\numberwithin{equation}{section}

\makeatother

\newcommand{\bea}{\begin{eqnarray}}
\newcommand{\eea}{\end{eqnarray}}
\newcommand{\be}{\begin{equation}}
\newcommand{\ee}{\end{equation}}

\newcommand{\eq}[1]{\begin{align}#1\end{align}}

\newcommand{\eqsp}[1]{\begin{equation}\begin{split}#1\end{split}\end{equation}}

\newcommand{\zbar}{\overline{z}}

\newcommand{\Z}{{\mathbb Z}}

\newcommand{\K}{{\mathsf{K}}}

\def\t{\tau}
\def\tb{\bar\tau}

\newcommand{\cC}{{\cal C }}

\newcommand{\cN}{{\cal N }}

\newcommand{\ie}{{i.e.,~}}

\newcommand{\zb}{{\bar z}}

\renewcommand{\title}[1]{\vbox{\center\LARGE{#1}}\vspace{5mm}}
\renewcommand{\author}[1]{\vbox{\center#1}\vspace{5mm}}
\newcommand{\address}[1]{\vbox{\center\footnotesize\em#1}}
\newcommand{\email}[1]{\vbox{\center\footnotesize\tt#1}\vspace{5mm}}

\begin{document}

\begin{titlepage}

 \begin{flushright}
 \hfill{\tt CALT-TH 2022-029}
\end{flushright}

\begin{center}

\hfill \\
\hfill \\
\vskip 1cm

 \title{The Stranger Things of\\
  Symmetric Product Orbifold CFTs}

\author{Nathan Benjamin$^{a}$, Suzanne Bintanja$^{b}$, Alejandra Castro$^{b}$, and Jildou Hollander$^b$
}

\address{
${}^a$Walter Burke Institute for Theoretical Physics \\ California Institute of Technology, Pasadena, CA 91125, USA
\\
${}^b$ Institute for Theoretical Physics, University of Amsterdam, Science Park 904, \\
1090 GL Amsterdam, The Netherlands
}
\email{nbenjami@caltech.edu, s.bintanja@uva.nl, a.castro@uva.nl, j.s.hollander@uva.nl}

\end{center}

\vfill

\abstract{Symmetric product orbifold theories are valuable due to their universal features at large $N$. Here we will demonstrate that they have features that are not as pervasive: we provide evidence of strange behaviour under deformations within their moduli space. To this end, we consider the symmetric product orbifold of tensor products of ${\cal N}=2$ super-Virasoro minimal models, and classify them according to two criteria. The first criterion is the existence of a single-trace twisted exactly marginal operator that triggers the deformation. The second criterion is a sparseness condition on the growth of light states in the elliptic genera. In this context we encounter a strange variety: theories that obey the first criterion but the second criterion falls into a Hagedorn-like growth. We explain why this may be counter-intuitive and discuss how it might be accounted for in conformal perturbation theory. We also find a new infinite class of theories that obey both criteria, which are necessary conditions for each moduli space to contain a supergravity point.
}

\vfill

\end{titlepage}

\eject

\tableofcontents
\section{Introduction}

The symmetric product orbifold theory of a two-dimensional conformal field theory (CFT) $\mathcal{C}$ is given by
\be\label{eq:sym}
{\rm Sym}^N({\cal C})\coloneqq \frac{\mathcal{C}^{\otimes N}}{S_N} \,.
\ee
That is, one tensors $N$ times a two-dimensional CFT $\cal C$, with central charge $c_0$, and orbifolds by the symmetric group $S_N$ permuting any of the $N$ theories. The resulting theory is a CFT with central charge $c=Nc_0$. The spectrum of such theories was computed in \cite{Dijkgraaf:1996xw}. Assuming that $\cal C$ is a compact, unitary CFT, the orbifold procedure by the symmetric group gives several appealing properties in the limit $N\to \infty$.

First, the orbifold has an effect analogous to gauging the theory: it keeps states that are invariant under $S_N$, the untwisted sector, and adds new states labelled by conjugacy classes of $S_N$, the twisted sectors. This assures that the correlation functions in ${\rm Sym}^N({\cal C})$ comply with large-$N$ factorization \cite{Pakman:2009zz,Pakman:2009ab,Belin:2014fna,Haehl:2014yla,Belin:2015hwa}. Second, there is a Hawking-Page phase transition at large $N$ \cite{Keller:2011xi,Hartman:2014oaa}. These are favorable features in the context of AdS/CFT.

There are also two additional properties that are key for their holographic interpretation: the universal Hagedorn growth of light states \cite{Keller:2011xi}, and the appearance of higher spin currents due to the orbifold structure, see e.g. \cite{Gaberdiel:2015uca,Apolo:2022fya}. Based on these two features, and concrete constructions in string theory, a ${\rm Sym}^N({\cal C})$ theory is usually interpreted as a weakly coupled CFT, whose dual can be matched to a tensionless limit of string theory \cite{Maldacena:1997re,Dijkgraaf:1998gf,Giveon:1998ns,Seiberg:1999xz,Argurio:2000tb,David:2002wn,Eberhardt:2018ouy,Giribet:2018ada,Eberhardt:2019ywk}. It is expected that under a suitable deformation, this weakly coupled system will reach a strongly coupled point, where the Hagedorn growth is dramatically lowered and the higher spin symmetry is gone. At strong coupling, the gravitational dual is expected to be well approximated by a semiclassical supergravity theory on AdS$_3$. These expectations are what we want to explore and  challenge here. 

Our aim here is to distinguish among different ${\rm Sym}^N({\cal C})$ theories based on their response under deformations. In particular, we want to turn on a marginal coupling in the theory: a coupling that preserves the large-$N$ factorization, and introduces interactions that modify the spectrum of the theory significantly. With this in mind, we will present criteria that allow us to infer responses under the deformations based on properties of the seed theory ${\cal C}$. Although we cannot prove or disprove the existence of a strongly coupled point even when the criteria are satisfied, we will see distinct patterns regarding the existence of a deformation and the response of ${\rm Sym}^N({\cal C})$ under it.  

Let us discuss more specifically how we plan to distinguish the different theories. We will classify symmetric product orbifolds based on properties of their moduli space. Here moduli space refers to the collection of exactly marginal operators that can be used to change the couplings of the theory, i.e., a conformal manifold. One can  establish very precisely the existence of moduli in supersymmetric theories, and therefore we will focus our work on superconformal field theories (SCFT) with at least ${\cal N}=2$. The possible operators that define the moduli space, and the behaviour of protected quantities therein, will give us two criteria to make distinctions among distinct Sym$^N({\cal C})$ theories.

The first criterion is the existence of an exactly marginal operator that breaks the orbifold structure of Sym$^N({\cal C})$: this way of introducing interactions should lift the higher spin currents. As reviewed recently in \cite{Belin:2020nmp,Apolo:2022fya}, it requires that the operator is single-trace, $\frac{1}{2}$-BPS, and in the twisted sector. This condition already places restrictions on ${\cal C}$ that we will review in the upcoming sections. More interestingly, we will see how it interplays with our second criterion.

The second criterion is related to the behaviour of the elliptic genera of Sym$^N({\cal C})$ theories at large $N$. The elliptic genus is an index that  counts $\frac{1}{4}$-BPS states weighted by $(-1)^F$; more importantly it is a protected quantity on the moduli space. This provides an extremely useful diagnostic: if the growth of light states is Hagedorn, i.e., exponential, for the elliptic genus,\footnote{See Sec.\,\ref{sec:criteria} and App.\,\ref{app:sparsenesscondition} for a more detailed definition of the growth conditions.} it immediately rules out any possibility of having a strongly coupled point with a spectrum that resembles a supergravity theory holographically. 

As we explore and classify Sym$^N({\cal C})$ theories based on these two criteria, we will encounter some fundamental limitations. The space of all unitary and compact SCFTs ${\cal C}$ is not known. Still there is an interesting subset of theories that one can systematically analyze: the ${\cal N}=2$ super-Virasoro minimal models. The work of \cite{Belin:2020nmp} showed that when  ${\cal C}$ is a single minimal model, the two criteria are satisfied: there is at least one single-trace, twisted, $\frac{1}{2}$-BPS marginal operator, and the growth of light states in the elliptic genus is sub-Hagedorn, i.e., slow. Here we will consider instead cases where ${\cal C}$ is a tensor product of ${\cal N}=2$ super-Virasoro minimal models. This situation is much richer and gives rise to strange and unexpected results. On the one hand we find a new infinite class of theories that obey both criteria, whose moduli spaces thus meet the necessary conditions to contain a supergravity point. We will however also see that both criteria are not automatically satisfied. In particular, we will find cases where we can find a suitable marginal deformation, but the elliptic genus displays Hagedorn-like behaviour. This challenges the lore on how to achieve a strongly coupled point that lifts the aforementioned properties of Sym$^N({\cal C})$. 

\subsection{Deformations of symmetric products}

It is useful to describe more concretely how we are envisioning a deformation of a symmetric product orbifold theory. In particular, we will describe the basic setup and how it affects some of the observables.  The idea is simple: we will turn on an marginal operator in Sym$^N({\cal C})$. From a path integral perspective, this means that we are adding to the action a term of the form
\eq{
\Delta S(\lambda)\coloneqq\lambda\sqrt{N}\int d^2z\ \Phi(z,\zbar)\,.\label{eq:defaction}
}
Here $\Phi$ is the deformation operator, with weights $(h,\bar h)=(1,1)$; in our case it is also a single-trace twisted sector operator that is exactly marginal, \ie its dimension is protected to all orders in perturbation theory. Furthermore, $\lambda$ is a coupling constant, and we have added a factor of $\sqrt{N}$ that ensures that there is a well-defined planar limit as $N\rightarrow\infty$ (with this convention $\lambda$ is independent of $N$). 

There are three observables that we will consider in this work. The first one is $\Phi$, and the question at hand is: can one establish the existence of at least one operator satisfying the appropriate requirements? The second observable is the elliptic genus. As mentioned above it is a protected quantity, \ie independent of $\lambda$. This is useful since it is an excellent diagnostic over the entire moduli space. In particular, it gives a lower bound on the growth of states that is insensitive to the value of $\lambda$.    

The third type of observables we will discuss are anomalous dimensions.\footnote{See, e.g., \cite{Avery:2010er,Gaberdiel:2015uca,Keller:2019yrr,Apolo:2022fya}, for more details. The following is a very short summary of the key concepts.} These are extracted from  correlation functions in the deformed theory, and their precise definition is as follows. A normalized two-point function in the deformed CFT is given by
\eq{
\left\langle O_a(z)O_b(z')\right\rangle_{\lambda}=\frac{\left\langle O_a(z)O_b(z')e^{\Delta S(\lambda)}\right\rangle}{\left\langle e^{\Delta S(\lambda)}\right\rangle}\,.\label{eq:defcor}
}
After deforming by a marginal operator, the resulting two-point-function is still constrained to be of the form:
\begin{align}
\left\langle \mathbb{O}_a(z) \mathbb{O}_b(z')\right\rangle_{\lambda}&=\frac{\delta_{ab}}{(z-z')^{2(h_a+\mu_{a}(\lambda))}(\bar z-\bar z')^{2(\bar h_a+\mu_{a}(\lambda))}}\,, \label{eq:deformation}
\end{align}
where $h_a$ and $\bar h_a$ are the left- and right-moving dimensions of the original, undeformed operator $\mathbb{O}_a$, and $\mu_a(\lambda)$ is the anomalous dimension. (Because spin is quantized, it cannot continuously vary, so the corrections to $h_a$ and $\bar h_a$ are forced to be the same.) As usual, the operators $\mathbb{O}_{a,b}$ are chosen to have a diagonal two-point function, so they are generally not the same as the operators $O_{a,b}$.

As we compare the different behaviour of the theories under study, we will report on the values of $\mu_{a}$ at leading order in $\lambda$.
 This first correction comes from expanding the exponentials in \eqref{eq:defcor} in orders of $\lambda$; evaluating the integrals in the expansion leads to an expression of the form
\eq{\label{eq:OOlambda}
\left\langle O_a(z)O_b(z')\right\rangle_{\lambda}=\frac{1}{(z-z')^{2h_a}(\zb-\zb')^{2\bar h_a}}\big(\delta_{ab}-2\gamma_{ab}\log\abs{z-z'}^2+ \ldots \big)\,.
}
The anomalous dimensions $\mu_a$ are then equal to the eigenvalues of the matrix $\gamma_{ab}$, while the eigenvectors of $\gamma_{ab}$ describe the operator mixing relating $O$ to $\mathbb{O}$. Anomalous dimensions can thus be found by computing the correction terms in \eqref{eq:defcor} and extracting the factor that multiplies the logarithm.

Being able to evaluate $\mu_a(\lambda)$ is key to understand what happens to the spectrum of Sym$^N({\cal C})$ once the deformation is turned on. In this work, we will denote a CFT to be {\it holographic} if for a finite value of $\lambda$ the vast majority of states get lifted (that is, very many become arbitrarily heavy, including the higher spin currents), and as a result the asymptotic growth of light states is significantly slowed relative to the fast Hagedorn growth. This is the ``very sparse'' condition, which requires that the CFT has very few operators of low scaling dimension \cite{Hartman:2014oaa,Belin:2016yll,Belin:2016dcu}. 

The tools of conformal perturbation theory are  still too rudimentary to give us enough control on $\mu_a(\lambda)$. For this reason we are taking more indirect routes via supersymmetric quantities, and we only discuss $\mu_a(\lambda)$ for holomorphic operators at leading order in $\lambda$. Despite these limitations, we find some interesting patterns that we will discuss in Sec.\,\ref{sec:contrast}. 

\subsection{Outline}
In this manuscript we study symmetric product orbifolds theories whose seed theory $\cal C$ corresponds to the tensor product of ${\cal N}=2$ super-Virasoro minimal models. In App. \ref{app:N2}, we define conventions of $\cN=2$ SCFTs and App. \ref{app:minimalmodels} discusses the $\cN=2$ super-Virasoro minimal models.
We classify these theories based on two criteria, discussed in Sec. \ref{sec:criteria}:
\begin{description}
 \setlength\itemsep{-0.2em}
    \item[Criterion 1:] Existence of single-trace twisted moduli. 
    \item[Criterion 2:] Slow growth of elliptic genera. 
\end{description}
The formula to check whether a theory contains such moduli is described in App. \ref{app:moduli}, the procedure to extract the growth of the elliptic genus is discussed in App. \ref{app:sparsenesscondition}.\footnote{It is worth noting that both criteria individually require $c_0 \leq 6$ \cite{Belin:2020nmp}.}
These give us four types of theories that we label from I-IV:
\begin{description}
\setlength\itemsep{-0.2em}
\item[Type I:] Theories that obey both criteria.
\item[Type II:] Theories that obey only criterion 1.
\item[Type III:] Theories that obey neither criteria.
\item[Type IV:] Theories that obey only criterion 2.
\end{description}

The classification of tensor products of minimal models into these four types is given in Sec. \ref{sec:tensormm}, where we find that they fall into the types I, II and III. To gain a better understanding of the different types, we contrast type II theories with type III theories in Sec \ref{sec:slope}. This is done by studying the parameter that regulates the Hagedorn-like growth. Moreover, we contrast type II theories with type I theories in Sec \ref{sec:deformation}, by studying the effect of the deformation on higher spin currents. Our results are summarised and discussed in Sec. \ref{sec:discussion}. In App. \ref{app:unicorns} we give a proof that modularity alone is not enough to rule out that type IV theories exist.

\section{Classification criteria}\label{sec:criteria}

Despite many shared properties of Sym$^N({\cal C})$, there are certain features that distinguish them. These are particularly important to signal if Sym$^N({\cal C})$ could lead to a holographic CFT. In this section, we will present the two properties used in this work to differentiate between different Sym$^N({\cal C})$ theories.   

Our construction is catered to superconformal theories in two dimensions with at least ${\cal N}=(2,2)$ supersymmetry and a discrete spectrum. This class of theories is interesting because the behaviour and anatomy of BPS quantities are highly dependent on the properties of the seed theory ${\cal C}$.  We will briefly review these quantities and their disparities, which will be the foundation of our classification in subsequent sections.  

\paragraph{Criterion 1: Existence of single-trace twisted moduli.} We denote as moduli the set of exactly marginal deformations. In the context of ${\cal N}=(2,2)$ theories, these correspond to certain $\frac{1}{2}$-BPS operators with appropriate charges and dimensions. For reasons reviewed recently in \cite{Apolo:2022fya}, it is as well crucial that they are single-trace and twisted operators. The existence of this class of operators depends on  the properties of ${\cal C}$: both its central charge and operator spectrum.   

A Sym$^N({\cal C})$ theory that satisfies this criterion 1 is a theory in which we can identify at least one $\frac{1}{2}$-BPS twisted sector operator that single-trace, and of the appropriate weight. The positive cases will be denoted as {\it moduli}. It is important to stress that we will only use this terminology for single-trace and twisted sector operators in moduli space; we will not report on other classes of exactly marginal deformations (e.g. operators that are multi-trace or untwisted states).  In this context, the term {\it no-moduli}  means the theory has no exactly marginal operators that are both single-trace and twisted. 

\paragraph{Criterion 2: Growth of elliptic genera.} For supersymmetric theories, the elliptic genus contains valuable information about the spectrum: it captures $\frac{1}{4}$-BPS states, weighted by $(-1)^F$, and it is independent of couplings.  As a result, we can check the growth of low-lying states at large $N$; if the growth is Hagedorn, then at no point in the moduli space can there be a weakly-coupled holographic theory \cite{Benjamin:2015hsa, Benjamin:2015vkc}. Following the results in \cite{Belin:2019rba,Belin:2019jqz,Keller:2020rwi}, we will use this criterion to determine if the growth of light states at large $N$ in the elliptic genera is compatible with a holographic CFT.

Along the lines of  criterion 2, we will denote if a Sym$^N({\cal C})$ theory as {\it slow} if the asymptotic growth of low-lying states at large $N$ is very sparse. We will denote a Sym$^N({\cal C})$ theory as {\it fast} if the asymptotic growth of low-lying states at large $N$ is Hagedorn-like.\footnote{The nomenclature follows the one introduced in  
\cite{Belin:2019rba,Belin:2019jqz}. If $d(\Delta)$ counts the low-lying states with conformal dimension $\Delta$ in Sym$^N({\cal C})$, {\it slow} means $\ln d(\Delta)\sim O(\Delta^\gamma)$, with $0\leq\gamma<1$, for $1\ll \Delta \ll N$. The term {\it fast} or {\it Hagedorn-like} means $\ln d(\Delta)= \nu \Delta + \dots $, with $0<\nu\leq 2\pi$, for the aforementioned range.} \\

With these two criteria, we will coin theories using the following four labels.
\begin{description}[leftmargin=0cm]
\item[Type I: slow and moduli.] A necessary, but not sufficient, condition for Sym$^N({\cal C})$ to be type I  is that the central charge of ${\cal C}$ is in the range $1\leq c_0\leq6$ \cite{Belin:2019jqz}. In fact, the unitary ${\cal N}=2$ super-Virasoro minimal models --- i.e., all unitary and compact SCFTs  with central charge in between $1\le c_0<3$ --- fall into this class \cite{Belin:2020nmp}. For $3\leq c_0\leq6$  these two requirements are very stringent, and relative to their cousins, there are very few and difficult to come by --- like a needle in haystack.  We hope that these type of theories will lead to a holographic CFT. 
\item[Type II: fast and moduli.] These theories are rare beasts. They will challenge the notion that the existence of single-trace twisted moduli is sufficient to argue for a strongly coupled point. They can occur in the range  $3\leq c_0\leq6$. We will find many examples here. 
\item[Type III: fast and no-moduli.] The most common occurrence in our classification. As shown in \cite{Benjamin:2015vkc,Belin:2019jqz}  (see also \cite{Belin:2020nmp}) any theory ${\cal C}$ with central charge $c_0>6$ will fall into this category. There are also plenty examples within the range $3\leq c_0\leq6$, which we use to contrast against the other types. 
\item[Type IV: slow and no-moduli.] The unicorns of the landscape. So far we have not found a unitary theory that complies with this behaviour. However, it is possible to design modular functions that are compatible with their existence. We will illustrate this in an example in App.\,\ref{app:unicorns}. To prove that these theories do not exist, modular invariance alone seems insufficient.

\end{description}

The exploration in the next section will focus on seeds theories in the range $3\leq c_0\leq6$, since it is the only window of ${\cC}$ where it is meaningful to discuss all four types of theories. One interesting finding, that we will continue to stress, is that we have not found a unitary and compact theory ${\cC}$ that is type IV. This indicates that criterion 2 might imply criterion 1, and hence is a stronger criteria to construct a holographic CFT. In other words, the slow growth in the elliptic genera might not be a coincidence, and should be taken as a key indicator that there is a point in moduli space where the theory is sparse and strongly coupled at large $N$.\footnote{As a technical point, we only consider theories whose elliptic genera do not vanish, and whose NS-sector vacua contribute in a nontrivial way to their elliptic genera.
Otherwise one could consider theories with fermion zero modes. 
 In some sense, theories with vanishing $Z_{\rm EG}$ trivially have ``slow" growth (though one could reconsider this analysis with a modified index \cite{Maldacena:1999bp}).
}

\section{Tensor products of minimal models}\label{sec:tensormm}

In this section we implement and classify SCFTs based on the criteria described in Sec.\,\ref{sec:criteria}. The interesting range that potentially could contain type I-IV theories is $3\leq c_0\leq6$, for the reasons mentioned before.


For seed theories of central charge $3\le c_0\le 6$ the situation is subtle, partially because there is no classification of them, and thus no exhaustive search that covers all SCFTs. Hence, we will proceed by looking at simple, but systematic,  constructions of SCFTs and establish their properties. In \cite{Belin:2020nmp} some examples of theories of both type I and III were found among the Kazama-Suzuki theories. Up to now there were no known CFTs of type II and IV. In this work we provide an exhaustive classification of tensor products of minimal models. 
They will divide themselves into type I, II and III theories (introduced in Sec.\,\ref{sec:criteria}). Tensor products of minimal models do not contain type IV theories. 

To start, we will quickly review some basic facts about minimal models; more information is presented in App.\,\ref{app:N2} and App.\,\ref{app:minimalmodels}. The $\cN=2$ super-Virasoro minimal models fully classify the unitary and compact $\cN=(2,2)$ SCFTs, with central charge less than 3 \cite{Boucher:1986bh,DiVecchia:1986fwg}. They are indexed by a positive integer $k$, that relates to the central charge via
\eq{
c_0=\frac{3k}{k+2}\label{eq:mmc}\,.
}
The minimal models are minimal in the sense that they consist of a finite number of irreducible representations of the $\cN=2$ superconformal algebra. The characters of the algebra must then combine into a modular invariant partition function, that additionally is invariant under spectral flow by a half unit. Possible ways to do this are in one-to-one correspondence with the ADE classification of simply laced Dynkin diagrams \cite{Gepner:1987qi,Cecotti:1992rm,Gray:2008je}. The partition function in the Ramond sector is given by
\eq{
Z^{\Phi}_{\rm RR}(\t,\tb,z,\zbar)=\frac{1}{2}\sum_{r,r'=1}^{k+1}N_{r,r'}^\Phi\sum_{s\in\Z/(2k+2)\Z}\tilde{\chi}^r_s(\t,z)\tilde{\chi}^{r'}_s(\tb,\zbar)\,.\label{eq:mmpf}
}
Here $\Phi$ is one of the simply laced Dynkin diagrams that are given by the $A$-, $D$-, and $E$-series. $N_{r,r'}^\Phi$ is the multiplicity matrix corresponding to this Dynkin diagram, and $\tilde{\chi}^r_s$ are characters related to the $\cN=2$ algebra.

The tensor product theories that we are interested in are given by
\eq{
\cC\coloneqq\bigotimes_{i=1}^{\K}\cC_{\Phi_i}\,,\label{eq:tp}
}
where with $\cC_{\Phi_i}$ we denote one of the $\cN=2$ super-Virasoro minimal models, and with $\K$ we denote the number of factors in the tensor product. The central charge of $\cC$ is:
\eq{
c_0=\sum_{i=1}^{\K}\frac{3k_i}{k_i+2}\,.\label{eq:tmmc}
}
The Ramond sector partition function of the tensor product theory is just the product of the partition functions of the individual minimal models:
\eq{
Z_{\rm RR}^\cC(\t,\tb,z,\zbar)=\prod_{i=1}^{\K}\ Z^{\Phi_i}_{\rm RR}(\t,\tb,z,\zbar)\,.\label{eq:tpf}
}

From the partition function of the tensor product, we can easily recover both the elliptic genus, by setting $\zbar=0\,$, and the $\frac{1}{2}$-BPS spectrum, by furthermore specializing to $q=\bar q=0\,$. These quantities are crucial to investigate the holographic criteria. The elliptic genera and $\frac{1}{2}$-BPS spectra of the $\cN=2$ super-Virasoro minimal models are given in App. \ref{app:minimalmodels}. Just as was the case for the Ramond sector partition function, the elliptic genus of the tensor product is then just given by the product of the elliptic genera of the minimal model factors in the tensor product, which are fractions of theta functions. Hence, to investigate criterion 2 we should check the sparseness condition, reviewed in App. \ref{app:sparsenesscondition}, for products of fractions of theta functions.

In order to study criterion 1 we must investigate the existence of single-trace twisted sector moduli in  the symmetric product of $\cC$. To ensure that the modulus is exactly marginal we require that it is a $G^\mp_{-1/2}$ descendant of a $\frac{1}{2}$-BPS primary of weight $h=\frac{1}{2}$ and $U(1)$ charge $Q=\pm1$. To have such a primary that is also single-trace and in the twist-$L$ sector, the following relation should be satisfied:
\eq{
\sum_{i=1}^{\K} \left(\frac{r_i}{k_i+2}-\frac{1}{2}\right) + \frac{c_0L}{6} \overset{!}{=}1\,.\label{eq:mod}
}
See App. \ref{app:moduli} for a derivation of this requirement. Here the $r_i$ are labels for the $\frac{1}{2}$-BPS states in the minimal model $\Phi_i$. The allowed values for the $r_i$ are given in App. \ref{app:minimalmodels}; with these conventions $r_i=1$ corresponds to the NS ground state. We will adopt the same notation for the twisted moduli as \cite{Belin:2020nmp}:
\eq{
(\underbrace{r_1...r_1}_{L \text{ times}}, ..., \underbrace{r_{\K}...r_{\K}}_{L \text{ times}})\,,
}
where the tensor product $\cC$ consist of $\K$ tensor factors of minimal models.

We are now ready to list our results. They will be organized in the following way: first by type (I through IV), and then by number of factors $\K$ in the tensor product. Furthermore, to avoid redundancies, we will not consider products that include the factors  $A_2  \otimes A_2$, $A_2 \otimes A_3$ and $A_2 \otimes A_4$, since
\eq{
A_2  \otimes A_2 = D_4\,, \qquad A_2 \otimes A_3 = E_6\,,\quad \text{and}\quad A_2 \otimes A_4 = E_8\,.\label{eq:redundancies}
} 
Lastly, we will consider tensor products with $\K \le 4$, since four tensor factors is the largest amount that does not include the aforementioned products and satisfies $c_0\le 6$.

\paragraph{Type I:}
These are the theories that contain single-trace twisted moduli and have slow growth in their elliptic genera. They pass both tests needed to have a weakly-coupled gravity dual, and are thus good candidates for potential theories dual to semiclassical supergravity (though evidently, these two tests are necessary but not necessarily sufficient).

\begin{description}[leftmargin=0cm]
\item[$\K=2$:]
The results have been summarised in Table \ref{tab:2typeI}. The tensor product of any two $A$-series theories is of type I. For tensor products involving $D$- or $E$-series, the classification is slightly more complicated. We also report on the single-trace and twisted moduli.
   
\item[$\K=3$:] 
There are only a few ways to construct tensor products of three minimal models with central charge $c_0 \le 6$. 
Besides a finite number of sporadic combinations, such as $A_2 \otimes A_6 \otimes X_{k \leq 39}$ and $A_4 \otimes A_4 \otimes X_{k \leq 7}$, where $X\in\{A,D,E\}$, there are three infinite families of theories with $c_0 < 6$ that can be characterised by
\be
A_2 \otimes A_5 \otimes X, \qquad A_2 \otimes D_4 \otimes X, \quad \text{and} \quad A_3 \otimes A_3 \otimes X\,.\label{eq:k3inf}
\ee
This is caused by the fact that  $A_2\otimes A_5$, $A_2 \otimes D_4$ and $A_3 \otimes A_3$ are the only ways to tensor two minimal models with total central charge $c_0 =3$.\footnote{We can also tensor product theories to make the total central charge $c_0 < 3$, but this will land us on another minimal model (see (\ref{eq:redundancies})), so we will not include it in our classification to avoid overcounting.} It turns out that the theories in \eqref{eq:k3inf} are of type I if and only if $X$ is an $A$-series minimal model. The finitely many sporadic combinations (that are not included in \eqref{eq:k3inf}) are of type III.

Furthermore, there are precisely 34 combinations of three minimal models with $c_0=6$. Interestingly enough, all of them are of type I, with one twist-2 modulus of the form $(11,11,11)$. 
The type I theories with $\K=3$ are summarized in Table \ref{tab:3typeI}.

\item[$\K=4$:] Subject to the redundacies in \eqref{eq:redundancies}, there is only one theory with $\K=4$: $(A_3)^4$, and it is of type I. The central charge is $c_0=6$ and the modulus is $(11,11,11,11)$, i.e., a twist-2 NS ground state. 
\end{description}

\paragraph{Type II:}
These are the theories which contain a single-trace twisted moduli, but do not exhibit slow growth in their elliptic genera. They are interesting to study, because even though they have a single-trace modulus, one cannot take this modulus to sufficiently strong coupling such that the Hagedorn density of states lifts to a supergravity-like growth.
\begin{description}[leftmargin=0cm]
\item[$\K=2$:]See Table \ref{tab:2typeII}. In fact these are the \emph{only} theories of type II that come from tensor products of minimal
models. 
\end{description}

\paragraph{Type III:}
These are the theories that have no single-trace twisted moduli and also no slow growth in their elliptic genera. 
\begin{description}[leftmargin=0cm]
\item[$\K=2$:] The results are given in Table \ref{tab:2typeIII}.
\item[$\K=3$:] The theories in this category are all tensor products of three minimal models that did not appear in Table \ref{tab:3typeI}, \ie the infinite families in \eqref{eq:k3inf} for $X$ a $D-$ or $E-$series minimal model, and a finite number of sporadic combinations with $c_0<6$.
\end{description}

 \begin{longtable}{|>{\centering\arraybackslash}p{3.5cm}|>{\centering\arraybackslash}p{3.5cm}|c|>{\centering\arraybackslash}p{3.5cm}|}
       \hline 
      Type I & $k_1$, $k_2$& Moduli &  Constituents\\ 
        \hline 
        \hline
\multirow{9}{3.5cm}{\centering{$A_{k_1+1}\otimes A_{k_2+1}$, $k_1 \leq k_2$}}
&$k_1 =1, ~ k_2 = 4$ & 1 twist-2, 1 twist-3 & (22,22), (111,111)\\
\cline{2-4}
&\multirow{2}{*}{$k_1 = k_2 = 2$} & \multirow{2}{*}{3 twist-2, 1 twist-3}& \multirow{2}{3.5cm}{\centering{(22,22), (11,33), (33,11), (111,111)}}\\
&&&\\
\cline{2-4} 
&\multirow{2}{*}{$k_2 = k_1 \geq 3$} & \multirow{2}{*}{3 twist-2}& \multirow{2}{3.5cm}{\centering{(22,22), (11,33), (33,11)}}\\
&&&\\
\cline{2-4} 
&\multirow{2}{3.5cm}{\centering{$k_2 = k_1 +n(k_1 + 2)$, $n \in \mathbb{Z}_{>0}$}}  &\multirow{2}{*}{2 twist-2}& \multirow{2}{3.5cm}{\centering{(22,22), (11;$(3+n,3+n)$)}}\\
& & & 
\\
\cline{2-4} 
& \multirow{2}{3.5cm}{\centering{$k_1 \neq k_2$ $\mod k_1+2 $}} &\multirow{2}{*}{1 twist-2} & \multirow{2}{*}{(22,22)}\\
&&&\\
\hline
\multirow{3}{3.5cm}{\centering{$D_{k_1/2+2}\otimes D_{k_2/2+2}$, $k_1 \leq k_2$}} & \multirow{2}{*}{$k_1 = k_2 = 4$} & \multirow{2}{*}{4 twist-2} & \multirow{2}{3.5cm}{\centering{(11,33), (33,11), (11,$\hat{3}\hat{3}$), ($\hat{3}\hat{3}$,11)}}\\
& & &\\
\cline{2-4} 
& $k_1 = k_2 \geq 6$ & 2 twist-2 & (11,33), (33,11)\\
\hline 
\multirow{7}{*}{$A_{k_1+1}\otimes D_{k_2/2+2}$} & \multirow{2}{*}{$k_1=1,~k_2=4$} & \multirow{2}{*}{2 twist-2, 1 twist-3} &\multirow{2}{3.5cm}{\centering{(11,33), (33,11), (111,111)}}\\
&&&\\
\cline{2-4} 
& \multirow{2}{*}{$k_1 = k_2=  4$} & \multirow{2}{*}{3 twist-2}& \multirow{2}{3.5cm}{\centering{(11,33), (33,11), (11,$\hat{3}\hat{3}$)}}\\
& & &\\
\cline{2-4} 
& $k_1 = 2$, $k_2 =6$ & 1 twist-2 & (11,44)\\
\cline{2-4} 
& $k_1 = 1$, $k_2 = 10$& 1 twist-2 & (11,66)\\
\cline{2-4} 
&$k_1 = k_2 \geq 6 $ & 2 twist-2 & (11,33), (33,11)\\
\hline 
\multirow{2}{*}{$A_{k_1+1} \otimes E_6$} & $k_1=2,~k_2=10$&1 twist-2&(11,55)\\
\cline{2-4}
& $k_1 = 4,~k_2=10$ & 1 twist-2 & (11,44)\\
\hline 
$A_{k_1+1} \otimes E_7$ & $k_1=4,~k_2=16$&1 twist-2&(11,55)\\
\hline
$A_{k_1+1} \otimes E_8$ & $k_1=4,~k_2=28$&1 twist-2&(11,77)\\
\hline 
\multirow{2}{*}{$D_{k_1/2+2}\otimes E_6$} &$k_1=4,~k_2=10$& 1 twist-2 & (11,55)\\
\cline{2-4}
& $k_1 = k_2 = 10$ & 1 twist-2 & (33,11)\\
\hline 
\multirow{2}{*}{$D_{k_1/2+2}\otimes E_7$} &$k_1=4,~k_2=16$& 1 twist-2 & (11,55)\\
\cline{2-4}
& $k_1 = k_2 = 16$ & 1 twist-2 & (33,11)\\
\hline 
\multirow{2}{*}{$D_{k_1/2+2}\otimes E_8$} &$k_1=4,~k_2=28$& 1 twist-2 & (11,77)\\
\cline{2-4}
& $k_1 = k_2 = 28$ & 1 twist-2 & (33,11)\\
\hline 
    \caption{Overview of the different single-trace twisted operators for sparse theories with two minimal models in the tensor product. Details on the notation can be found in App. \ref{app:moduli}.}
    \label{tab:2typeI}
\end{longtable}
\newpage
  \begin{longtable}{|>{\centering\arraybackslash}p{3.8cm}|c|>{\centering\arraybackslash}p{6cm}|}
    \hline
       Type I & Moduli & Constituents \\
       \hline \hline
        \centering{$A_2\otimes (A_5)^2$}  & 2 twist-2 & (11,11,22), (11,22,11)\\
       \hline
               \centering{$(A_3)^3$}  & 3 twist-2  & (11,11,22), (11,22,11), (22,11,11)\\
       \hline
        \centering{$A_2\otimes A_5\otimes A_{k+1}$}  & 1 twist-2 & (11,11,22)\\
       \hline
              \centering{$A_2\otimes D_4\otimes A_{k+1}$}  &1 twist-2 & (11,11,22)\\
              \hline
      \centering{$(A_3)^2\otimes A_{k+1}$}  & 1 twist-2  & (11,11,22)\\
\hline 
\multirow{2}{3.8cm}{\centering{Any combination with $c_0 = 6$}} & \multirow{2}{*}{1 twist-2} & \multirow{2}{*}{(11,11,11)}\\
& & \\
\hline 
\caption{Overview of the different single-trace twisted operators for sparse theories with three minimal models in the tensor product, with $c_0<6$. Here we have $k\in\mathbb{Z}_{>0}$.}
    \label{tab:3typeI}
\end{longtable}

\begin{longtable}{|>{\centering\arraybackslash}p{3.5cm}|>{\centering\arraybackslash}p{5cm}|c|>{\centering\arraybackslash}p{3.8cm}|}
  \hline
\def \arraystretch{1.3}
  Type II & $k_1$, $k_2$ & Moduli & Constituents \\
       \hline \hline
       \multirow{2}{3.5cm}{\centering{$D_{k_1/2+2}\otimes D_{k_2/2+2}$, $k_1 \leq k_2$}} & \multirow{2}{5cm}{\centering{$k_2 = k_1 + 2n(k_1+2)$}} &\multirow{2}{*}{1 twist-2} & \multirow{2}{*}{(11;$(3+2n,3+2n)$)}\\
       & & & \\
       \hline
       \multirow{2}{3.5cm}{\centering{$A_{k_1+1}\otimes D_{k_2/2+2}$}} & $k_2 = k_1 + 2n(k_1+2)$ &1 twist-2& (11;($3+2n,3+2n$))\\
       \cline{2-4}
      & $k_1 = k_2 + n(k_2+2)$ & 1 twist-2 & $((3+n,3+n);11)$\\
       \hline
$A_{k_1+1} \otimes E_i$ & $k_1 = k_{2,i} + (n-1)(k_{2,i}+2)$ &1 twist-2&$((2+n,2+n);11)$\\
\hline 
$D_{k_1/2+2} \otimes E_i$ & $k_1 = k_{2,i}+ 2n(k_{2,i}+2)$ & 1 twist-2 & ($(3+2n,3+2n)$;11)\\
\hline 
    \caption{Overview of the different single-trace twisted operators for non-sparse theories. For $i=6,7,8$ one has $k_{2,i}=10,16,28$. In all cases, $n \in \mathbb{Z}_{>0}$.}
    \label{tab:2typeII}
\end{longtable}

    \begin{longtable}{|>{\centering\arraybackslash}p{3.5cm}|>{\centering\arraybackslash}p{6cm}|}
    \hline 
        Type III & $k_1,~k_2$  \\
        \hline \hline
   \multirow{2}{3.5cm}{\centering{$D_{k_1/2+2}\otimes D_{k_2/2+2}$, $k_1 \leq k_2$}}  & \multirow{2}{6cm}{\centering{$k_2 \neq k_1 \mod 2(k_1+2)$}}\\
    & \\
   \hline 
  \multirow{2}{*}{$A_{k_1+1}\otimes D_{k_2/2+2}$} & \multirow{2}{5cm}{\centering{$k_1 \neq k_2 \mod 2(k_1+2)~$ and $k_1 \neq k_2 \mod k_2+2$}}\\
  &\\
  \hline
  $A_{k_1+1} \otimes E_i$ & $k_1 \neq k_{2,i} \mod k_{2,i}+2$\\
  \hline
  $D_{k_1/2+2} \otimes E_i$ & $k_1 \neq k_{2,i} \mod 2(k_{2,i}+2)$\\
  \hline 
    \caption{Theories which have neither slow growth nor single-trace twisted operators with two minimal models in the tensor product. For $i=6,7,8$ one has $k_{2,i}=10,16,28$. }
    \label{tab:2typeIII}
\end{longtable}

\paragraph{Type IV:}
Type IV theories have slow growth in their elliptic genera, but no moduli. 
There are no tensor products of minimal models that fall into this category. 

\FloatBarrier
\section{Contrasting theories}\label{sec:contrast}

In this section we will try to create some order in the chaos that is the classification of the tensor products of minimal models into type I, II, and III. In particular, we look for patterns that distinguish between the different types of theories. 
Our main goal is to understand type II theories, in contrast to theories of type I and III. Recall that type II theories are theories that do contain a modulus, but their elliptic genus exhibits Hagedorn-like (fast) growth. Therefore, their moduli space cannot contain a holographic point, in the sense that we will never meet the very sparse condition on the spectrum.

It is important to stress that the mere existence of type II theories is somewhat counter-intuitive. If there is a modulus, the lore is that all operators that are not protected should get lifted under the deformations. This has been explored in \cite{Gaberdiel:2015uca,Keller:2019suk,Guo:2020gxm,Benjamin:2021zkn,Apolo:2022fya,Guo:2022ifr} using conformal perturbation theory, 
where one consistently gets non-trivial anomalous dimension $\mu_a$ for the operators considered. Type II theories are evidence that there is tension in this understanding of the conformal manifold: the elliptic genus of type II is telling us that too many states are unaffected as $\lambda$ is turned on, but the intuition from conformal perturbation theory would say this is unlikely. In the following we will attempt to find a signal in the seed theory ${\cal C}$ that could resolve this conflict.

From our analysis in Sec.\,\ref{sec:tensormm}, type II theories only occur when $\K=2$. That is, when the seed theory is of the form
\be
{\cal C}=X_{k_1}\otimes Y_{k_2}~,
\ee
with $X,Y \in\{A,D,E\}$. There are two notable features behind this structure. As displayed in Table \ref{tab:2typeII}, criterion 1 for type II is satisfied due to having a suitable operator belonging to $X$ ($Y$); the other portion of the marginal deformation sits on the twisted NS ground state of $Y$ ($X$). On the other hand, the fast growth in criteria 2 relies on the combined spectrum of $\frac{1}{4}$-BPS states from both $X$ and $Y$ --- recall that $X$ and $Y$ are individually slow. These features motivate how we will contrast and scrutinize type II theories. 

In Sec.\,\ref{sec:slope}, we investigate the differences in the elliptic genus between theories of type II and III. In both these classes the elliptic genus exhibits fast growth, despite one of the classes having a modulus and the other one not. In an attempt to have a more refined view of what it means to be ``fast,'' we will compare the magnitude of the parameter that controls the exponential growth for fixed central charge of the seed, $c_0$. We will find that type III theories are not necessarily faster than type II, although they ``usually" are. In particular, if the operator content between the theories is very similar, type II theories are slower than their type III counterpart. We will compare a type II theory $X_{k_1}\otimes Y_{k_2}$ versus a type III theory with $X_{k_2}\otimes Y_{k_1}$. 

In Sec.\,\ref{sec:deformation}, we contrast type I and II theories by investigating the effect that the marginal operator has on the anomalous dimensions $\mu_a$ of higher spin currents, by using conformal perturbation theory. The aim here is to see if the composition of the marginal operator is significant. We will study in detail an example where we fix $c_0$, the composition of the marginal operator is exactly the same for a type I and II theory, and $Y_{k_2}$ is shared too; we deem this as a case where the two types are the most similar. Unfortunately, we have not detected any differences on $\mu_a$ at leading order for unprotected currents. We will gain some insight on what should happen for generic operators.

\subsection{Contrasting fast growth}\label{sec:slope}
For type II theories, the elliptic genus exhibits fast growth, despite the existence of an exactly marginal deformation along the lines of criterion 1. 
This means that turning on the marginal deformation for these theories should have a mild effect on the spectrum. In particular, it will not reduce the Hagedorn-like (fast) growth of light operators to a supergravity-like (slow) growth.  Still, this mild effect might be reflected on a reduced parameter that controls the fast growth in the elliptic genus. This potential reduction is what we will explore in the following, by making suitable comparisons between type II and type III theories.   

To make the comparison more precise, we need to quantify the asymptotic growth of these light states.
The slope of the Hagedorn-like growth in the elliptic genus is regulated by the  value of $\tilde{\alpha}$, defined in App. \ref{app:sparsenesscondition}.
If $\tilde{\alpha}$ is positive, the elliptic genus of $\operatorname{Sym}^N(\cC)$ has Hagedorn-like growth with parameter $\nu$ in the large-$N$ limit \cite{Belin:2019rba}:
\be
d(\Delta) \sim \exp\left(2\pi \sqrt{\frac{24\tilde{\alpha}}{c_0}}\Delta\right) = \exp (\nu \Delta)~, \qquad 1 \ll \Delta \ll N\,,
\ee
where
\be
\nu \coloneqq 2\pi \sqrt{\frac{24\tilde{\alpha}}{c_0}}\label{eq:nualpha}\,.
\ee
Both the partition function of $\operatorname{Sym}^N(\cC)$ and the unsigned count of $\frac14$-BPS states of $\operatorname{Sym}^N(\cC)$ universally have this type of growth with $\nu=2\pi$ \cite{Keller:2011xi, Benjamin:2016pil}. Since the elliptic genus is the index of $\frac{1}{4}$-BPS states, the maximal value that $\nu$ can have is $2\pi$. In the following we will report on the value of $\nu$, or alternatively $\tilde\alpha$, for type II and III theories with $\K=2$. 

\begin{figure}
\begin{center}
    \begin{minipage}{0.28\linewidth}
    \includegraphics[scale=0.58]{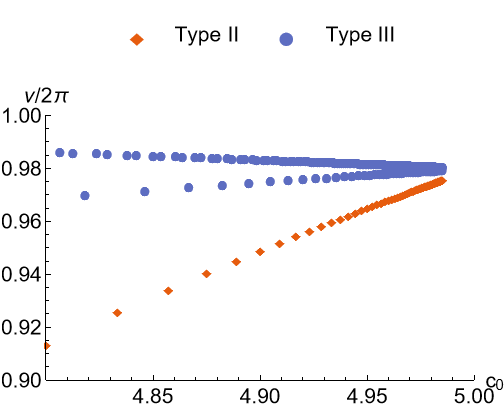}
    \subcaption{$A_{k+1} \otimes D_4$}
    \label{fig:AiD4}
    \end{minipage}
    \hspace{0.04\linewidth}
    \begin{minipage}{0.28\linewidth}
    \includegraphics[scale=0.58]{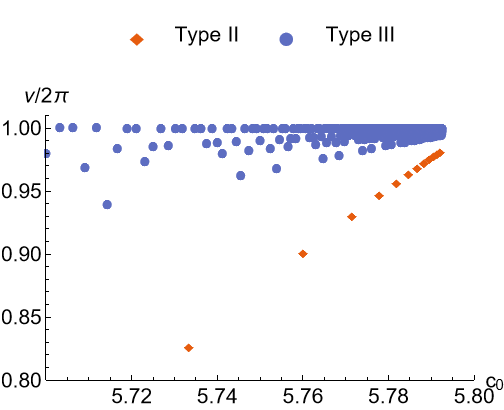}
    \subcaption{$D_{k/2+2} \otimes E_8$}
    \label{fig:DiE8}
    \end{minipage}
    \hspace{0.04\linewidth}
     \begin{minipage}{0.28\linewidth}
    \includegraphics[scale=0.58]{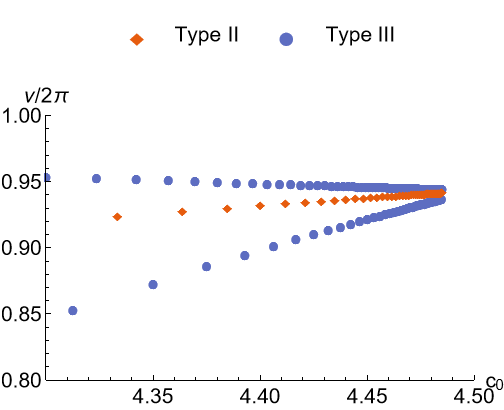}
    \subcaption{$A_3 \otimes D_{k/2+2}$}
    \label{fig:A3Di}
    \end{minipage}
    \caption{Comparison of the largest value of $\nu/2\pi$ as a function of the central charge for $A_{k+1}\otimes D_4$ (\ref{fig:AiD4}), $D_{k/2+2}\otimes E_8$ (\ref{fig:DiE8}) and $A_3 \otimes D_{k/2+2}$ (\ref{fig:A3Di}). The maximal value $\nu/2\pi$ can assume is 1, corresponding to exactly Hagedorn growth. Note that in some case $\nu$ approaches $2\pi$, but never reaches it. 
    }
    \label{fig:TypeIIvsIII}
\end{center}
\end{figure}

In Fig. \ref{fig:TypeIIvsIII}, the value of $\nu/2\pi$ is shown for theories of the form $A_{k+1}\otimes D_4$, $D_{k/2+2}\otimes E_8$ and $A_3 \otimes D_{k/2+2}$ for different values of $k$. In all three cases, the different theories seem to follow separate trajectories. Fig.\,\ref{fig:AiD4} and \ref{fig:DiE8} seem to suggest that the trajectory of $\nu/2\pi$ for type II theories, with moduli, lies below the trajectory for type III theories, without a modulus. However, in Fig.\,\ref{fig:A3Di}, the trajectory for type II theories lies between two trajectories for type III, so something more intricate is happening. 

There is one claim we can prove easily with an example: at fixed $c_0$, a type II theory does not necessarily have a smaller value of $\nu$ than a type III theory. 
A simple example is the following: compare $A_7 \otimes D_{21}$ and $A_{14}\otimes D_7$. Both have $c_0=\frac{51}{10}$, but the type II theory has faster Hagedorn growth 
than the type III theory, since
\be
\begin{aligned}\label{eq:counter-example}
A_7 \otimes D_{21}~~(\text{type II}): &\qquad \tilde{\alpha} =\frac{9}{51}\,, \quad \nu = \frac{8\pi}{17}\sqrt{15}\,,\\
A_{14}\otimes D_7~~ (\text{type III}): &\qquad \tilde{\alpha} = \frac{8}{51}\,, \quad \nu = \frac{16\pi}{17} \sqrt{\frac{10}{3}}\,.
\end{aligned}
\ee

Nevertheless, Fig.\,\ref{fig:TypeIIvsIII} does seem to suggest there is a pattern between the existence of the modulus and a reduced slope of the Hagedorn-like growth. 
 Given the example in \eqref{eq:counter-example}, one has to consider a more restricted setting than just fixing $c_0$. Since we are looking at tensor products of minimal models, one might try to compare theories with the same values for the central charges of the separate factors in the tensor product.
More precisely, we will consider the following conjecture:\\

\textit{Let $X_{k_1}$ and $Y_{k_2}$ be minimal models parameterized by $k_1$ and $k_2$ respectively. If the symmetric product of  $X_{k_1}\otimes Y_{k_2}$ contains an exactly marginal single-trace twisted operator and $X_{k_2}\otimes Y_{k_1}$ does not, then $X_{k_1}\otimes Y_{k_2}$ has a smaller slope of the Hagedorn growth than $X_{k_2}\otimes Y_{k_1}$.}\\

The only types of theories where such asymmetry in the moduli can occur is $A_{k_1+1}\otimes D_{k_2/2+2}$. More specifically, when $k_1+2 = 2n(k_2+2)$ (where $n \in \mathbb{Z}_{>0}$), $A_{k_1+1}\otimes D_{k_2/2+2}$ contains an exactly marginal single-trace twisted operator, while $A_{k_2+1}\otimes D_{k_1/2+2}$ does not. Therefore, the former is a type II theory while the latter is a type III theory. By computing $\tilde{\alpha}$ for $n\leq 25$ and $k_2\leq 500$, the theories follow
\begin{equation}
\begin{aligned}
\tilde{\alpha}(n,k_2) &= \frac{k_2(2n-1)}{4(2n(k_2+1)-1)}~, \qquad  \text{type II}, \quad k_2 = 0 \mod 4\,,\\
\tilde{\alpha}(n,k_2) &= \frac{(k_2-2)(2n-1)}{4(2n(k_2+1)-1)}~, \qquad \text{type II}, \quad k_2 = 2 \mod 4\,,\\
\tilde{\alpha}(n,k_2) &= \frac{nk_2-2}{2(2n(k_2+1)-1)}~, \qquad \text{type III}\,.
\label{alphamax}
\end{aligned}
\end{equation}
One can show that the states that cause these values of $\tilde{\alpha}$ occur in the spectrum for all $n$ and $k_2$.
 We have checked them explicitly for as many as 6200 theories, providing significant confidence in their correctness for any value of $n$ and $k_2$.\footnote{However, a rigorous proof or derivation of \eqref{alphamax} has so far been elusive.}
From \eqref{alphamax}, it is then straightforward to show that
\be
\tilde{\alpha}_{\text{type II}}(n,k_2) \leq \tilde{\alpha}_{\text{type III}}(n,k_2)\,,
\ee
where this inequality is saturated if and only if $k_2 = 4$. Therefore, we can say with a high level of confidence that our conjecture 
is true. This supports the expectation that the existence of the modulus lowers the growth of the low-lying states in the elliptic genus, albeit in a restricted setting.

\subsection{Effects of the marginal deformations}\label{sec:deformation}

Next, we will attempt to understand type II theories based on the composition of their moduli, and study the effects on the spectrum. In particular, we investigate the change in dimension of some simple operators. In this way, we can compare type II theories to type I theories. Which operators we consider, and why the computations are simple will be explained shortly.


From Sec.\,\ref{sec:tensormm} we notice a few things about type II theories. First of all, the only theories of type II consist only of two tensor factors (\ie $\K=2$ for all type II theories). Second, all the moduli that appear in type II theories have twist equal to two.\footnote{This is because of the range of $c_0$: for seed theories with central charge in the range $3<c_0\leq6$, one can show that the only twisted sector moduli that can appear are those of twist-2 \cite{Apolo:2022fya}.} This fact simplifies the computations needed to evaluate $\mu_a$ perturbatively. Third, from Table \ref{tab:2typeII} we see a pattern in the constituents of the moduli of type II theories; it is constructed out of states of the form 
\eq{\label{eq:modK2}
(11,rr)\,.
}
The fact that one of the states has label $r_i=1$ gives a particular simple structure of the modulus; this will alleviate some of the technicalities  as we will explain below.  There are also a few type I theories with this same composition for the moduli, as illustrated in Table\,\ref{tab:2typeI}. Our aim is to compare the effect of deformation of the form \eqref{eq:modK2} on certain unprotected operators  for type I versus type II.

Let us elaborate on the precise form of the moduli. For cases with $\K=2$ we will have\footnote{We denote with $\hat\Phi$ only the chiral part of the full Hermitian modulus $\Phi$ to keep notation simple. The computations and logic is completely analogous when using the full $\Phi$. Recall that $\Phi$ is the exactly marginal operator with $(h,\bar h)=(1,1)$.} 
\eq{
\hat\Phi=\left(G_{-1/2}^-\otimes\mathbb{1}+\mathbb{1}\otimes G_{-1/2}^-\right)\left(\varphi_{\Phi_1}\otimes\varphi_{\Phi_2}\right)\sigma_2\times\text{anti-holomorphic}\,.\label{eq:mod1}
}
Our conventions for the supercurrents $G^\pm$ are in  App.\,\ref{app:N2}. Here $(\varphi_{\Phi_{1}}\otimes\varphi_{\Phi_{2}})\sigma_2$ is a $\frac{1}{2}$-BPS primary of weight $h=1/2$ in the twist-2 sector. That is, it is an excitation of the twist-2 ground state $\sigma_2$. For all type II theories  (and also plenty of type I theories) one of the tensor factors of this operator is the twist-2 ground state, \ie $\varphi_{\Phi_i}$ is a ground state for $i=1$ or $i=2$. The fact that $\varphi_{\Phi_i}$ is a ground state is directly linked to the label $r_i$ being equal to one.

Next, let us discuss the type of operators we will consider. We want to evaluate \eqref{eq:OOlambda}. For a general operator $O_a$, albeit straightforward conceptually, it is quite complicated to extract $\gamma_{ab}$. The basic reason is that one is dealing with at least a four-point function. The appearance of twisted states in \eqref{eq:mod1} also complicates the process. Therefore, to keep the technicalities at a minimum, we will consider primary operators $O_a$ that are in the \textit{untwisted} sector of Sym$^N({\cal C})$ and also \textit{holomorphic}.  

These holomorphic operators are clearly conserved currents. They are interesting because they are generically not protected by symmetry along the conformal manifold, but they do lead to significant simplifications when evaluating \eqref{eq:mod1}. Moreover, in order to have a low-energy effective field theory description of supergravity, there cannot be any higher spin currents in the spectrum. Therefore, checking that any higher spin current that is present at the orbifold point will lift  (or not) is key to establish a holographic point on the moduli space. 

Symmetric product orbifold theories contain several classes of currents, with arbitrary high spin; some are more universal and some depend on the details of the seed theory. We can split these currents into three families:\footnote{Note that we can also have a higher spin current that are a mixture of different families. For example, we can have a seed current on one tensor factor, with a tensor current on another. This does not impact our conclusions.}
\begin{description}[leftmargin=0.2cm]
\item[Symmetric product currents.] These are currents constructed out of the $\cN=2$ algebra currents of the seed theory, and are present purely because of the structure of the symmetric product. The construction and lifting of the first few symmetric product currents is investigated in \cite{Apolo:2022fya}, where it is shown that they lift for all seed theories that belong to type I or II.

\item[Tensor currents.] These are currents come about because the seed theory we consider is a tensor product. We can use the tensor product to construct new currents out of the $\cN=2$ algebra of the individual tensor factors. For example, in case of the stress tensor $T$, we can use
\eq{
T\otimes\mathbb{1}\,,\qquad\text{and}\qquad \mathbb{1}\otimes T\,.
}
We should compare this to the symmetric product currents, that are constructed only using the stress tensor of the tensor product
\eq{
T_\text{seed}=T\otimes\mathbb{1}+\mathbb{1}\otimes T\,.
}
Because the structure of tensor currents is not much different than symmetric products, we expect (and verify for specific cases) that all currents will lift along the lines of \cite{Apolo:2022fya}.

\item[Seed currents.]We will denote seed currents of spin $s$ with $\mathcal{W}_s$. They are constructed using enhanced symmetries of the seed theory. In our setup they originate from the $D$- and $E$-series minimal models, which have an enlarged chiral algebra compared to the $\cN=2$ algebra, see App. \ref{app:minimalmodels} for an overview, that can be used to construct currents.
\end{description}

In the following we will discuss the leading effect in $\lambda$ for the anomalous dimensions $\mu_a$ of these currents. We will place emphasis on the seed currents, since they fall outside the results in  \cite{Apolo:2022fya}. We will find  that all currents considered lift, and hence there are no signals of a difference between type I and II theories. We will illustrate this by considering a particular simple example, after which we end the section with some general remarks.

\subsubsection{Concrete example: \texorpdfstring{$E_6\otimes X$}{E6xX}}

To be concrete, in this subsection we will consider seed theories we consider are of the form
\eq{
{\cal C}=E_6\otimes X\,.\label{eq:E6X}
}
Here we pick $X$ such that the theory is type I or II, and such that the modulus is labeled by
\eq{
(11,rr)\,,
}
\ie the part of the modulus that sits on $E_6$ is a twisted ground state. For example, if $X=A_{11}$, the theory is type II and has modulus $(11,33)$, whereas $X=D_7$ gives a type I theory and also has modulus $(11,33)$. The contrast between these two choices will illustrate sharply the odd behaviour of type II theories. We have theories of the same central charge, same values of $k_i$, and with moduli that look exactly the same, yet one is of type I, while the other is of type II. 

The current we consider is of the following form. As noted in App. \ref{app:minimalmodels}, $E_6$ contains an additional spin-1 current (next to the $U(1)$ current of the $\cN=2$ algebra) that we will denote with $J_1$ \cite{DiVecchia:1986fwg,Gray:2008je,Cappelli:1987xt}. We can use $J_1$ to construct the following current in the symmetric products of \eqref{eq:E6X}
\eq{
\mathcal{W}_1\coloneqq \frac{1}{\sqrt{N}}\sum_{i=1}^N J_1^{(i)}\otimes \mathbb{1}^{(i)}\,,\label{eq:currentdef}
}
where the upper index $i$ denotes the different sheets of the orbifold theory. Note that we have defined the current such that all its weight is on the $E_6$ factor of the seed. Together with the fact that the modulus is such that it has no nontrivial weight on $E_6$, this makes the setup in some sense optimal to study the interplay with the modulus \eqref{eq:mod1}. Basically, this is one of the simplest states one might suspect is not keen to interact with $\Phi$.  

To compute the lifting of $\mathcal{W}_1$ we must compute the correction terms in \eqref{eq:defcor} with $O_{a,b}=\mathcal{W}_1$. The first nonzero correction term appears at order $\lambda^2$ and is given by
\eq{
\frac{\lambda^2N}{2}\int d^2w \ d^2w' \ \left\langle  \mathcal{W}_1(z) \Phi(w)\Phi(w')\mathcal{W}_1(z')\right\rangle_{\text{connected}}\label{eq:lsq}\, .
}
The fact that there is no linear term can be understood from unitarity. If there was one, one could choose the sign of $\lambda$ to obtain a negative anti-holomorphic weight, which is not allowed in a unitary theory.

To extract the anomalous dimension we can make use of the fact that the operator that we are computing the lifting of is a current. This fact can be used to perform the anti-holomorphic integrals using Ward identities, which will give rise to a logarithm, whose coefficient determines the anomalous dimension \cite{Gaberdiel:2015uca,Apolo:2022fya}. Once all the dust is settled, the matrix elements of the lifting matrix are determined by contour integrals around the operator insertions
\eqsp{
\gamma_{\mathcal{W}_1\mathcal{W}_1} &=  \frac{ \lambda^2  N (z - z')^{2}  }{16}\bigg(\oint_z d w \oint_{z'} d w' + \oint_{z'} d w \oint_z d w'\bigg)\\
&\hspace{120pt}\times(\bar w-\bar w')^{2} \left\langle  \mathcal{W}_1(z) \tilde{\Phi}(w)\tilde{\Phi}(w')\mathcal{W}_1(z')\right\rangle_{\text{connected}}\,. \label{gammadef}
}
Here we denote with  $\tilde\Phi$ the left-moving part of the full modulus $\Phi$.
It thus follows that the anomalous dimensions are completely determined at leading order by the connected contribution to the holomorphic four-point function with two insertions of the deformation operator.

In order to evaluate the four-point function in \eqref{gammadef}, it is simplest to first use Ward identities to exchange the $G_{-1/2}^\pm$ modes on the moduli for derivatives (see for example App. D of \cite{deBoer:2008ss}). To compute the resulting twisted correlator, we go to the $L$-fold cover, where $L$ denotes the twist of the moduli, using the so-called cover map \cite{Dixon:1986qv,Lunin:2000yv,Lunin:2001pw,Calabrese:2009qy,Pakman:2009zz,Pakman:2009ab}. The reason why one would use an $L$-fold cover to compute twisted sector correlators, is that twist operators get mapped to ordinary primaries or descendants of the seed theory on the cover space. Under the cover map, operators on the $i$-th tensor factor of the symmetric product get mapped to the $i$-th sheet on the cover, while twisted operators get mapped to branch points of the cover space. The correlator we must compute contains two twist-$L$ insertions at $w$ and $w'$ (for us $L=2$). For such a correlator, the cover map is given by
\eq{
\frac{x - w'}{x - w} = \frac{(t - t_{w'})^L}{(t - t_w)^L}\,, \label{cmap}
}
 where $x$ is the coordinate we use on the base space of the symmetric product, and $t$ is the coordinate we use on the cover space. Moreover, the twist insertions get mapped to insertions at $t_{w'}$ and $t_w$ on the cover. 
 NS sector twist operators of the form $\varphi\sigma_2$ of weight $1/2$ (which is the operator applicable to us) will get sent to a Ramond sector primary $\tilde\varphi$ on the cover of weight $\tilde h$, which satisfies
\eq{
\tilde h-\frac{c_0}{24}=1-\frac{c_0}{6}\,.
}
For the theories we consider, $\tilde\varphi$ takes a particularly simple form: on one of the tensor factors it is trivial. More precisely, whenever $r_i=1$ appears in the form of the modulus of a seed theory, its image under the cover map will be a Ramond ground state.

For the current defined in \eqref{eq:currentdef}, the correlator we must compute on the cover space then factorizes over the tensor product to \footnote{Here we keep prefactors of $N$, but we do not write any other factors multiplying the correlator.}$^,$\footnote{Had we not chosen the currents and moduli so carefully, this factorization would generally not hold, and there would be additional contributions to the four-point function in \eqref{gammadef}.}
\eqsp{
&\frac{1}{N}\left\langle J_1(t_z)J_1(t_{z'})\right\rangle_{E_6}\left\langle \tilde\varphi_{X}(t_w)\tilde\varphi_{X}(t_{w'})\right\rangle_{X}\\
&=\frac{1}{N}\frac{1}{(t_z-t_{z'})^2}\frac{1}{(t_w-t_{w'})^{2\tilde h_{r}}}\,,\label{eq:factorization}
}
where $\tilde h_r-\frac{c_X}{24}=1-\frac{c_0}{6}$. This step is key. The correlator factorizes on the cover due to the tensor product nature of ${\cal C}$, which gives two independent pieces---a piece controlled by the operator sitting on $E_6$ and a piece by the primary on $X$ that defines the modulus. Still, this contributes to the connected correlator \eqref{gammadef}. Combining the steps described above, we can finally evaluate the anomalous dimension of $\mathcal{W}_1$. We find that at leading order in $\lambda$
\eq{
\mu_{s=1}=\pi^2\lambda^2\,.
}
Here the matrix element $\gamma_{\mathcal{W}_1\mathcal{W}_1}$ is equal to the eigenvalue $\mu_{s=1}$ since at spin one there is (at most) one unique current (beside the $U(1)$ current of the $\cN=2$ algebra, which is protected from getting an anomalous dimension), and hence the lifting matrix is just a number.

\subsubsection{Remarks and generalizations}

We end this section with some remarks and generalizations of the concrete example we just gave. 

The first remark is that the factorization in \eqref{eq:factorization}, due to the composition of the modulus, did not spoil the lore coming from conformal perturbation theory. In particular, we obtained a non-trivial value for $\mu_{s=1}$, and we expect this to persist for other seed currents. 

Let us investigate this more carefully. Consider cases with $\K=2$ and the modulus of the form $(11,rr)$. For any spin-$s$ current of the form \eqref{eq:currentdef}, with $J_1$ replaced by a spin-$s$ current, our analysis in Sec.\,\ref{sec:deformation} follows in a straightforward way. In Table \ref{tab:gg} we give the matrix element $\gamma_{\mathcal{W}_s\mathcal{W}_s}$ of the lifting matrix for the spin-$s$ current $\mathcal{W}_s$.

\begin{table}[ht!]
	\centering
	\def\arraystretch{1.8}
	\begin{tabular}{|c|c|c|c|c|c|c|c|}
	\hline
		$s$  & 1  & 2 & 3 & 4& 5 & 6 & 7\\
		\hline
		$\frac{\gamma_{\mathcal{W}_s\mathcal{W}_s}}{\pi^2\lambda^2}$ & 1 & $\frac{9}{4}$  & $\frac{15}{4}$ & $\frac{175}{32}$ & $\frac{945}{128}$ & $\frac{4851}{512}$ & $\frac{3003}{256}$ \\
		\hline
	\end{tabular}
	\caption{Matrix elements $\gamma_{\mathcal{W}_s\mathcal{W}_s}$ for the first few values of $s$.}
	\label{tab:gg}
\end{table}

Remarkably, the matrix elements are very robust, and only depend on the existence of the current and modulus of the form such that we have factorization \eqref{eq:factorization}; there is no explicit dependence on the central charges of the minimal models in the tensor product. This is different from, e.g, the symmetric product currents; there the answers depend on $c_0$ and the detailed composition of the modulus \cite{Apolo:2022fya}.  

Note that Table \ref{tab:gg} contains matrix elements, and not full anomalous dimensions. The anomalous dimensions are given by the eigenvalues of the full mixing matrix. Therefore, these matrix elements cannot guarantee that the eigenvalues of the lifting matrix are nonzero. A miracle could occur where the lifting matrix becomes degenerate and has a zero eigenvalue. We have shown however, that such a miracle does not occur at spin two, when we consider all symmetric product and tensor currents, and seed currents that live on the tensor factor where the modulus has no nontrivial weight.

We reiterate that the factorization \eqref{eq:factorization} was crucial to simplify the computation and get our results. Nevertheless, we believe that the ultimate conclusion, namely that currents get lifted, does not depend on the factorization. This belief is rooted in the fact that even when we do not have factorization as in \eqref{eq:factorization}, there is always a term of the form \eqref{eq:factorization} where the four-point function splits into the product of two two-point functions. In this sense, our setup is minimal: there is only one term in the correlation function that contributes. It is true however, that this ``universal" contribution is already enough to generate a nonzero logarithmic term, or equivalently a nonzero $\gamma_{ab}$. We do not believe that any additional terms in the correlator on the cover space can spoil the existence of the logarithmic term, even though we have not checked this explicitly. 

For the reasons explained above we come to the following hypothesis:\\

\textit{All currents (not part of the $\cN=2$ algebra) get lifted in type I and type II theories.}\\

We thus conclude that currents are no good for distinguishing between type I and type II theories. Nonetheless, there must be a big difference between the spectra of type I and type II theories under deformation by the moduli, since the spectrum grows significantly slower in type I theories. The question remains however how we can see that difference explicitly. From the results here, and \cite{Apolo:2022fya}, it seems that to access differences in the lifting of the spectrum, one needs to consider the non-holomorphic spectrum. One concrete thing one could do to expose a difference between type I and II theories is to look at the $\frac{1}{4}$-BPS spectrum, where from the elliptic genus we know that for type I theories a large part of the spectrum must lift, while for type II theories not many states can lift.

\section{Discussion}
\label{sec:discussion}

In this paper we studied a family of symmetric orbifold CFTs -- namely symmetric orbifolds of tensor products of $\mathcal{N}=2$ minimal models -- and classified them based on two criteria: (1) the existence of single-trace marginal operators and (2) the growth of the density of states of the elliptic genus. Both the existence of a single-trace marginal operator and slow growth of the index are necessary (but not sufficient) conditions for the theory to have a weakly-curved Einstein gravity dual in the moduli space. 

The most striking results of our work are:
\begin{enumerate}[label=(\roman*)]
    \item We found a new infinite family of type I theories. These theories obey both criteria, which are necessary conditions for each moduli space to potentially contain a supergravity point. 
    \item We found theories of type II (obeying criterion 1 but not 2). This clearly shows that criterion 1 alone is not sufficient to find a holographic CFT. 
    \item We found no theories of type IV (obeying criterion 2 but not 1). This hints that criterion 2 is more stringent that criterion 1. It might also be possible that criterion 2 implies criterion 1, and hence it should be viewed as more fundamental.
\end{enumerate}

There are various interesting future directions that one could pursue, some of which we collate here.

\vspace{2mm}

\noindent\emph{More on type IV theories}

\vspace{2mm}

If a type IV theory existed, it would be the result of a ``mathematical accident" in the following sense. Since the theory has no marginal operators to move far from the symmetric orbifold, the density of states has to be Hagedorn everywhere in the moduli space \cite{Keller:2011xi}. Nonetheless these theories have a sub-Hagedorn growth in the index. This would mean there is an accidental cancellation that is not the result of the index being forced to match the slowest-growing point in moduli space. 

Of course, such mathematical accidents are not logically impossible. However, in the class of examples we have considered (tensor products of minimal models as seed theories) we have shown they are impossible. It would be very interesting if we could prove they never happen (which would mean criterion 2 is strictly stronger than criterion 1), or to find an example of a type IV theory. Note that modular invariance of the supersymmetric index alone is not enough to prove that type IV theories cannot exist (see App. \ref{app:unicorns}).

\vspace{2mm}

\noindent\emph{Lifting $\frac{1}{4}$-BPS states}

\vspace{2mm}

Can we quantify more precisely which and how many $\frac{1}{4}$-BPS states get lifted as one moves away from the orbifold point? For theories of type I, we expect the vast majority of states to be lifted (meaning, the asymptotic growth changes from exponential to sub-exponential). In this paper we began analyzing this at the level of the currents, but on general grounds we also expect a similar statement to be true for all BPS states. This was also analyzed for $K3$ in \cite{Keller:2019suk} and $\text{Sym}^N(\mathbb{T}^4)$ in e.g. \cite{Avery:2010er, Guo:2019ady, Benjamin:2021zkn, Guo:2020gxm, Guo:2022ifr}. 
Unfortunately it seems that the conformal perturbation theory technology may be too complicated to go to extremely large energies and track the BPS states, but this should in principle be possible to do. It may be especially interesting to do this analysis for the type II theories. For these theories, we expect some states to get lifted, but ``not many" of them (since there is an exponential growth in the index). If we can track carefully which states get lifted and which do not, it may give a more intuitive explanation as to why most do not get lifted. 

An even more challenging question would be to move towards calculating the anomalous dimension at finite coupling (as opposed to perturbatively close to the orbifold point). It is possible, for instance, that some states have anomalous dimensions that plateau to a certain value at large coupling. It would be interesting to see if type I and type II theories behave differently in this respect.

\vspace{6mm}

\noindent\emph{Other families of theories}

\vspace{2mm}

In this paper we have just analyzed a very special class of $\mathcal{N}=2$ theories with $3\leq c_0 \leq 6$, the tensor products of minimal models. Of course there is no classification of $\mathcal{N}=2$ theories with $c_0\geq 3$ so there is no a priori reason to expect the theories we studied in this range are generic. It would be nice if we could somehow do a more exhaustive search. 

More broadly, so far we have only been analyzing symmetric product orbifold theories. It would be interesting also to analyze other classes of large-$N$ theories and see where they fit in terms of criteron 1 and 2. See for example \cite{Gaberdiel:2010pz,Candu:2012jq,Candu:2012tr,Gaberdiel:2016xwo} for a class of theories with $W_N$ symmetry with ${\cal N}=2$ supersymmetry. Could we use similar techniques as the ones in this paper to study theories of higher spin gravity?

\section*{Acknowledgements}

We thank Ahmed Almheiri, Luis Apolo, Alex Belin, Shamit Kachru, and Christoph Keller for interesting discussions, and collaborations on related topics.
The work of SB, AC and JH is supported by the Delta ITP consortium, a program of the Netherlands Organisation for Scientific Research (NWO) that is funded by the Dutch Ministry of Education, Culture and Science (OCW). The work of NB is supported by the Sherman Fairchild Foundation and the U.S. Department of Energy, Office of Science, Office of High Energy Physics Award Number DE-SC0011632 .

\appendix

\section{Conventions on \texorpdfstring{${\cal N}=2$}{N=2} SCFTs}\label{app:N2}

In this appendix we collect some basic properties of unitary and compact $\cN=2$ superconformal theories, in particular their algebra and elliptic genera. Our conventions follow those in, e.g., \cite{Lerche:1989uy,Blumenhagen:2009zz}. 

The $\cN=2$ superconformal algebra contains a stress tensor $T$,  a weight-1 $U(1)$ current $J$, and two weight-3/2 fermionic currents $G^+$ and $G^{-}$ with $U(1)$ charges $+1$ and $-1$, respectively. In terms of its generators, the corresponding commutators are given by
\begin{equation}
\begin{aligned}\label{superVir comms}
\bigl[L_m,L_n\bigr]
 ={}&
  (m-n)L_{m+n} +\frac{c}{12}m(m^2-1)\delta_{m+n,0}\,,
   \\
\bigl[J_m,J_n\bigr]
={}&
\frac{c}{3}m\,\delta_{m+n,0}\,,
  \\
\bigl\{G^+_r,G^-_s\bigr\}
 ={}&
2L_{r+s}+(r-s)J_{r+s} +\frac{c}{3}\Big(r^2-\frac{1}{4}\Big)\delta_{r+s,0}\,,
 \\
\bigl[L_m,J_n\bigr]
={}&
 -n\, J_{m+n}\,, \\ 
 \bigl[J_m,G^\pm_r\bigr]
  ={}&
  \pm G^\pm_{m+r}\,,
 \\
  \bigl[L_m,G^\pm_r\bigr]
  ={}&
   \Big(\frac{m}{2}-r\Big) G^\pm_{m+r}\,,
   \\
\bigl\{G^\pm_r,G^\pm_s\bigr\}
={}&
0\ .
\end{aligned}
\end{equation}
Here the central charge is $c$ and $U(1)_R$ level is $\hat t=c/6$. The $\cN=2$ superconformal algebra is invariant under the  spectral flow automorphism \cite{Schwimmer:1986mf}:
\begin{equation}
\begin{aligned}\label{spectral flow 1}
L_n &\quad \to \quad\,\,\,\, L'_{n} ={} L_n + \eta J_{n} + \frac{\eta^{2}}{6}c \,\delta_{n,0}\,,
\\
J_n &\quad \to  \quad \,\,\,\, J'_{n}  ={} J_n + \frac{c}{3}\eta\,  \delta_{n,0}\,,
\\
G^{\pm}_{r} &\quad \to \quad  G_{r}^{\pm '}  ={} G^{\pm}_{r\pm \eta}\,.
\end{aligned}
\end{equation}
For $\eta \in \mathbb{Z} + 1/2$ the flow interpolates between the NS and R sectors, while for $\eta \in \mathbb{Z}$ it maps the NS and R sectors to themselves.

The representations of this algebra are parametrized by their weight $h$ and their $U(1)_R$ charge $Q$. 
There are two types of representations: long (or non-BPS) representations, and short (or BPS) representations. In the NS sector, the BPS states are of the form
\begin{equation} \label{eq:stateNS}
\left |\, h=\frac{\abs{Q}}{2}~, ~Q\right\rangle_{\rm NS}~,
\end{equation}
and depending on the sign of $Q$, we call it a chiral $(c)$ or anti-chiral $(a)$ field.  In the Ramond sector, this implies that 
\begin{equation} \label{eq:stateR}
\left |\,h=\frac{c}{24}~, ~Q\,\right\rangle_{\rm R}~,
\end{equation}
and we therefore call it a Ramond ground state.
Each state has a left- and right-moving component; if both components are BPS, we say the state is $\frac{1}{2}$-BPS, and if only one is BPS, we say it is $\frac{1}{4}$-BPS. 

One of the central objects we will consider is the elliptic genus
\eq{\label{ellipticgenus1}
Z_{\text{EG}}(\tau,z) = \text{Tr}_{\text{RR}}\ ((-1)^Fq^{L_0-\frac{c}{24}} y^{J_0} \bar{q}^{\bar{L}_0-\frac{c}{24}}\ )~, \qquad q\coloneqq e^{2\pi i\tau}~, ~~y\coloneqq e^{2\pi i z}\,,
}
which captures the $\frac{1}{4}$-BPS states in the Ramond sector. If the CFT has a discrete spectrum, then $Z_{\rm EG}$ is a holomorphic function of $\tau$, as it only receives contributions from right-moving ground states.

\section{Aspects of \texorpdfstring{${\cal N}=2$}{N=2} minimal models}\label{app:minimalmodels}

$\mathcal{N}=2$ minimal models are unitary SCFTs, and constitute a complete classification of unitary CFTs with $\mathcal{N}=(2,2)$ supersymmetry and central charge $c<3$ \cite{Boucher:1986bh, DiVecchia:1986fwg}. The minimal models are parametrized by a positive integer $k$, that relates to the central charge as
\be\label{eq:cmm}
c = \frac{3k}{k+2}~.
\ee
The classification of minimal models is naturally related to simply-laced Dynkin diagrams, which define the $A$-, $D$-, and $E$-series. These families are labeled as:
\begin{itemize}
\item The $A$-series, with $c=\frac{3k}{k+2}$ for any $k \in \mathbb{Z}_{>0}$ which we will call $A_{k+1}$. Besides the $\cN=2$ currents they contain no other currents. 
\item The $D$-series, with $c=\frac{3k}{k+2}$ for any even $k \geq 4$ which we will call $D_{k/2+2}$. On top of the $\cN=2$ currents, $D_{k/2+2}$ contains a current of spin $s=\frac{k}{4}$.
\item Finally there are three exceptional theories which we call $E_6$, $E_7$, and $E_8$, with $c=\frac 52, \frac 83$, and $\frac{14}5$, respectively. Each of the exceptional theories contain additional higher spin currents, besides the $\cN=2$ currents: $E_6$ contains an additional spin-1 current, $E_7$ contains an additional spin-4 current, and $E_8$ contains a spin-1, spin-3, and spin-7 current.
\end{itemize}
In the following we will summarize some main results about these theories, following the conventions stated in \cite{Belin:2020nmp}.

The highest weight representations in the Ramond sector are given by the following weights and $U(1)$ charges
\begin{align}
\label{eq:QR}
h_{r,s}^\epsilon=\frac{r^2-s^2}{4(k+2)}+\frac{c}{24}~ ,\qquad  Q_s^\epsilon=\frac{s}{k+2}+\frac{\epsilon}{2}\,.
\end{align}
The labels run as 
\begin{align}
r\in \{1, \ldots  k+1\}\,,\qquad \ 0\leq |s+\epsilon|\leq r-1\,, \qquad
\ r+ s = 0~ \text{mod}~ 2\,,
\end{align}
and $\epsilon=\pm1$; the positive integer $k$ labels the central charge of the minimal model. Through spectral flow these can be related to NS sector representations of weight and $U(1)$ charge
\begin{align}
h_{r,s}^\epsilon=\frac{r^2+2s-s^2+k}{4(k+2)}+\frac{\epsilon}{4}\,,\qquad Q_s^\epsilon=\frac{2s+k}{2(k+2)}+\frac{\epsilon}{2}\,.\label{eq:hns}
\end{align}

The $\frac{1}{2}$-BPS representations in the Ramond sector for an ADE theory correspond to states with $r=|s|$. The $\frac{1}{2}$-BPS partition function is given by 
\be\label{eq:modrmin}
Z^\Phi_{\frac{1}{2}-{\rm BPS}}=  \frac{1}{2} \sum_{0 <r<k+2}N^\Phi_{r,r}\left((y\bar y)^{\frac r{k+2} -\frac12}+ (y\bar y)^{-\frac r{k+2} +\frac12}\right)\, .
\ee
Here $N^\Phi_{r,r'}$ is Capelli-Itzykson-Zuber (CIZ) matrix \cite{Cappelli:1987xt}, and $\Phi$ is the simply-laced Dynkin diagram.  We note that the $\frac{1}{2}$-BPS partition function only depends on the diagonal entries of the CIZ matrix and is always diagonal, even for the $D$- and $E$-series models. 
By writing out the CIZ matrix one can find the $\frac{1}{2}$-BPS partition functions for the individual minimal models:
\paragraph{$A$-series.}
For $\Phi= A_{k+1}$, we find the following expression for the $\frac{1}{2}$-BPS partition function:
\be\label{eq:AMMhodge}
Z^{A_{k+1}}_{\frac{1}{2}-{\rm BPS}}(y, \bar y) = \sum_{r=1}^{k+1} \left (y\bar y\right)^{\frac{r} {k+2}-\frac{1}{2}}\,.
\ee
Moreover their elliptic genera are given by \cite{Witten:1993jg, DiFrancesco:1993dg}
\be
Z_{\text{EG}}^{A_{k+1}}(\tau, z) = \frac{\theta_1\left(\tau, \frac{k+1}{k+2} z\right)}{\theta_1(\tau, \frac z{k+2})}.
\ee

\paragraph{$D$-series.}
For $\Phi=D_{{k}/{2}+2}$, we find
\be\label{eq:DMMhodge}
Z^{D_{{k}/{2}+2}}_{\frac{1}{2}-{\rm BPS}}(y, \bar y) = 1+\sum_{j=1}^{\frac{k}{2}+1} \left (y\bar y\right)^{\frac{2j-1}{k+2}-\frac{1}{2}}\,,
\ee
where we can compare to the $A$-series be setting $r=2j-1$. Their elliptic genera are given by
\be
Z_{\text{EG}}^{D_{k/2+1}} = \frac{\theta_1\left(\tau, \frac{kz}{k+2}\right) \theta_1\left(\tau,\frac{k+4}{2k+4} z\right)}{\theta_1\left(\tau, \frac{2z}{k+2}\right) \theta_1\left(\frac{kz}{2(k+2)}\right)}\,.
\ee

\paragraph{$E$-series.}
For the $E$-type minimal models, we find
\begin{equation}
\begin{aligned}\label{eq:EMMhodge}
Z^{E_6}_{\frac{1}{2}{\rm -BPS}}(y, \bar y) =&(y\bar y)^{-\frac{5}{12}}+(y\bar y)^{-\frac{1}{6}}+(y\bar y)^{-\frac{1}{12}}+(y\bar y)^{\frac{1}{12}}+(y\bar y)^{\frac{1}{6}}+(y\bar y)^{\frac{5}{12}}\,,\\
Z^{E_7}_{\frac{1}{2}-{\rm BPS}}(y, \bar y) =&(y\bar y)^{-\frac{4}{9}}+(y\bar y)^{-\frac{2}{9}}+(y\bar y)^{-\frac{1}{9}}+1+(y\bar y)^{\frac{1}{9}}+(y\bar y)^{\frac{2}{9}}+(y\bar y)^{\frac{4}{9}}\,,\\
Z^{E_8}_{\frac{1}{2}-{\rm BPS}}(y, \bar y) =&(y\bar y)^{-\frac{7}{15}}+(y\bar y)^{-\frac{4}{15}}+(y\bar y)^{-\frac{2}{15}}+(y\bar y)^{-\frac{1}{15}}+(y\bar y)^{\frac{1}{15}}+(y\bar y)^{\frac{2}{15}}\\ &+(y\bar y)^{\frac{4}{15}}+(y\bar y)^{\frac{7}{15}}\,,
\end{aligned}
\end{equation}
and elliptic genera
\begin{equation}
\begin{aligned}    
Z_{\text{EG}}^{E_6}(\tau,z) &= \frac{\theta_1(\tau, \frac{3z}4)\theta_1(\tau, \frac{2z}3)}{\theta_1(\tau, \frac z4) \theta_1(\tau, \frac z3)}\,, \\
    Z_{\text{EG}}^{E_7}(\tau,z) &= \frac{\theta_1(\tau, \frac{7z}9)\theta_1(\tau,  \frac{2z}3)}{\theta_1(\tau,\frac{2z}{9})\theta_1(\tau,\frac{z}{3})} \,, \\
    Z_{\text{EG}}^{E_8}(\tau, z) &= \frac{\theta_1(\tau,\frac{4z}{5})\theta_1(\tau,\frac{2z}{3})}{\theta_1(\tau,\frac{z}{5})\theta_1(\tau,\frac z3)}\,.
\end{aligned}
\end{equation}

\section{Single-trace twisted moduli}\label{app:moduli}

One of the criteria that we will check is the existence of single-trace twisted moduli. In the NS sector, the moduli that can deform the CFT must be $G_{-1/2}^\pm$ descendants of (anti-) chiral primaries (or equivalently, $\frac{1}{2}$-BPS states) with $h = 1/2$ and $Q = \pm 1$ \cite{Belin:2020nmp}. 
In the Ramond sector, (anti-)chiral primaries correspond to ground states. In the seed theory, these are captured by
\be
    Z_{\frac{1}{2}-\text{BPS}}(y,\overline{y}) = \sum_{Q,\overline{Q}}d(Q,\overline{Q}) y^Q \overline{y}^{\overline{Q}}\,.
\ee
The generating function for the symmetric product in the Ramond sector then takes the following form \cite{deBoer:1998us}:
\be
    \sum_{N=0}^\infty Z_{\frac{1}{2}-\text{BPS}}^N (y,\overline{y}) p^N = \prod_{L=1}^\infty \prod_{Q,\overline{Q}}\frac{1}{(1-p^L y^Q \overline{y}^{\overline{Q}})^{d(Q,\overline{Q})}}~.
\ee
Here $Z_{\frac{1}{2}-\text{BPS}}^N (y,\overline{y})$ is the $\frac{1}{2}$-BPS spectrum of the symmetric product orbifold with $N$ copies and $L$ corresponds to the twist of the operator.
Focusing on the $(c,c)$ primaries, we can write down the generating function in the NS sector as
\be
    \sum_{N=0}^\infty Z_{cc}^N (y,\overline{y}) p^N = \prod_{L=1}^\infty \prod_{Q,\overline{Q}}\frac{1}{(1-p^L y^{Q+c_0L/6} \overline{y}^{\overline{Q}+c_0L/6})^{d(Q,\overline{Q})}}\,.
    \label{generating function in NS sector}
\ee
A single-trace $\frac{1}{2}$-BPS operator in $\operatorname{Sym}^N(\cC)$ of twist $L$ thus has a charge
\be 
\begin{aligned}
Q = Q_i + \frac{c_0 L }{6}\,,\\
\overline{Q} = \overline{Q}_i + \frac{c_0 L}{6}\,,
\end{aligned}
\ee
where $(Q_i,\overline{Q}_i)$ is the charge of one of the Ramond ground states in the seed. For $(c,c)$ primaries, we thus need to find $(Q_i,\overline{Q}_i)$ such that $(Q,\overline{Q}) = (1,1)$. 

In the Ramond sector, the ground states in the seed theory for the minimal models parametrized by $k$ have a charge
\be 
Q_r = \overline{Q}_r = \frac{r}{k+2} - \frac{1}{2}~,
\ee
where the range of $r$ depends on the type of minimal model we are considering: see \eqref{eq:AMMhodge}-\eqref{eq:EMMhodge}. 
We are looking for chiral primaries of $\cC = \bigotimes_{i=1}^{\K} \cC_i$, where $\cC_i$ are $\cN = 2$ super-Virasoro minimal models parametrized by $k_i$.
A state $\phi \in \cC$ that is built up as $\phi = \bigotimes_{i=1}^{\K} \phi_i, ~\phi_i \in \cC_i$ is a chiral primary of $\cC$ if and only if $\phi_i$ is a chiral primary in $\cC_i$ for all $i$.
Therefore, we are looking for solutions to
\be 
\begin{aligned}
Q = \sum_{i=1}^{\K} \left(\frac{r_i}{k_i+2}-\frac{1}{2}\right) + \frac{c_0 L}{6} \stackrel{!}{=}1~,
\end{aligned}
\ee
with $L\geq 2$. Since the central charge of $\cC$ is the sum of the central charges of the different $\cC_i$, we have
\begin{equation}
    \begin{aligned}
    c_0 = \sum_{i=1}^{\K} \frac{3k_i}{k_i +2}~.
    \end{aligned}
\end{equation}
We will use a similar notation to denote the different solutions of the moduli as the authors of \cite{Belin:2020nmp}:
\be(\underbrace{r_1...r_1}_{L \text{ times}}, ..., \underbrace{r_{\K}...r_{\K}}_{L \text{ times}}).\ee
Almost all charges are non-degenerate. However, if $\frac{k}{2}+1$ is odd, the charge with this value of $r$ appears twice in the $\frac{1}{2}$-BPS spectrum of $D_{k/2+2}$. To denote this degeneracy, we will write a hat if this special value happens to be odd.

\section{Sparseness condition on the elliptic genus}\label{app:sparsenesscondition}

In this appendix, we review the criterion first discussed in \cite{Belin:2019rba, Belin:2019jqz} in order to give necessary and sufficient conditions for a family of symmetric product orbifold CFTs to have slow-growing (sub-Hagedorn) elliptic genus at large $N$. The condition is on the elliptic genus of the seed theory. This criterion was used in \cite{Belin:2020nmp} for a large family of seed theories and we extend it in this paper. A more convenient formulation for a specific family of seed elliptic genera (in terms of Jacobi theta functions) was recently given in \cite{Keller:2020rwi}.

Let us consider a seed theory $\mathcal{C}$ that has $\mathcal{N}=(2,2)$ supersymmetry and elliptic genus $Z_{\text{EG}}^{\mathcal C}(\tau,z)$ (defined in \eqref{ellipticgenus1}). The ``standard" convention in the literature for the normalization of the $U(1)_R$ current is so that the supercharge of the $\mathcal{N}=2$ algebra has charge $\pm 1$. Under this convention, not all charges of operators are integers. (For example in $\mathcal{N}=2$ minimal models, under this convention, most operators have fractional charges.) We will instead normalize the $U(1)$ current so that all operators have integer charges with gcd $=1$. In \cite{Belin:2019rba, Belin:2019jqz}, this was called \emph{unwrapping} the elliptic genus. To be more precise, suppose under the convention where the supercharges have charge $\pm 1$, all operators have charge with gcd $\frac1\kappa$ for some integer $\kappa$. We define the unwrapped elliptic genus as
\be
\varphi(\tau, z) \coloneqq Z_{\text{EG}}(\tau, \kappa z) = \sum_{\substack{n\geq 0\\ \ell \in \mathbb{Z}}}c(n,\ell) q^n y^\ell\,.
\label{eq:zegwjf}
\ee
By definition $\varphi(\tau,z)$ now has an integer charge expansion and is a weak Jacobi form of weight $0$ and index $t$, with 
\be
t = \frac{c_0}{6} \kappa^2\,.
\ee
If we assume that the most polar term (which comes from the spectral flow of the NS vacuum) does not vanish in the elliptic genus, it is of the form $y^{c_0\kappa/6} q^0$. In the notation of \cite{Belin:2019rba,Belin:2019jqz,Belin:2020nmp} this was denoted by $b$ (i.e. $b=c_0\kappa/6$). With this notation out of the way, the criterion is then given by the following. For every nonvanishing term $q^n y^\ell$ in the unwrapped elliptic genus, compute the following quantity:
\be\label{alphacriterion2}
\alpha(n,\ell) \coloneqq \max_j
\left(-\frac{ t}{b^2}j\left(j-\frac{b \ell}{t}\right)-n\right)\,~~~~j=0, 1, \ldots, b-1\,.
\ee
In \eqref{alphacriterion2}, for a fixed $n,\ell$, we maximize the quantity in parenthesis over $b$ different values of $j$. If, for any term, $\alpha > 0$, then the NS-sector elliptic genus of the symmetric orbifold of $\mathcal{C}$ has Hagedorn growth at large $N$ (in the regime $1 \ll \Delta \ll N$). However, if for all coefficients in the elliptic genus, \eqref{alphacriterion2} has $\alpha \leq 0$, then the elliptic genus of the symmetric orbifold grows sub-Hagedorn, or supergravity-like. It was also shown that if $4tn-\ell^2 \geq 0$ (which defines a ``nonpolar" state), we have $\alpha(n, \ell) < 0$ automatically. Thus the criterion requires only a finite calculation to check (since there are only finitely many polar states with $4tn-\ell^2 < 0$).

In \cite{Keller:2020rwi}, a computationally simpler criterion was written down for seed elliptic genera of the following special form
\be\label{thetafunctionproduct}
\varphi(\tau, z) = \frac{\theta_1(\tau, n_1 z)\theta_1(\tau,n_2 z) \cdots \theta_1(\tau,n_\mathsf{N} z)}{\theta_1(\tau, m_1 z)\theta_1(\tau,m_2 z) \cdots \theta_1(\tau,m_\mathsf{N} z)}\,,
\ee
for $n_i, m_i \in \mathbb{Z}_{>0}$. If the NS vacuum contributes to the elliptic genus, the most polar term is of the form $q^0 y^{b}$ with $b = \frac{1}{2}\sum_{i=1}^\mathsf{N} (n_i-m_i)$. The authors of \cite{Keller:2020rwi} showed that we should compute the value of 
\be\label{alphacriterion3}
\begin{aligned}
\tilde\alpha &\coloneqq -\min_{j} \bigg(\sum_{i=1}^\mathsf{N} \frac{n_i^2-m_i^2}{2}\frac{j^2}{b^2}-\frac{n_i-m_i}{2}\frac{j}{b}+\frac{\lfloor \frac{n_i j}{b}\rfloor \left(\lfloor \frac{n_i j}{b}\rfloor + 1\right)}{2}\\
&-\frac{\lfloor \frac{m_i j}{b}\rfloor \left(\lfloor \frac{m_i j}{b}\rfloor + 1\right)}{2} - \frac{n_ij}{b}\left \lfloor \frac{n_i j}{b} \right \rfloor +  \frac{m_ij}{b}\left \lfloor \frac{m_i j}{b} \right \rfloor \bigg) \qquad j = 0,1,\ldots,b-1\,.
\end{aligned}
\ee
It can be shown that $\tilde \alpha$ and $\alpha(n,\ell)$ agree when $\alpha(n,\ell)$ takes its maximal value
\eq{
\tilde{\alpha}=\max_{n,\ell}\left(\alpha(n,\ell)\right)\,.
} 
If $\tilde{\alpha}>0$, the NS sector elliptic genus of the symmetric product orbifold has Hagedorn growth at large $N$. Moreover, the growth is parametrized by $\tilde{\alpha}$ in the following way:
\be
d(\Delta) \sim \exp\left(2\pi \sqrt{\frac{24 \tilde{\alpha} }{c_0}}\Delta\right) \qquad 1 \ll \Delta \ll N\,.
\ee
Conversely, if $\tilde{\alpha} \leq 0$, the coefficients of the NS sector elliptic genus of $\operatorname{Sym}^N(\cC)$ grow as
\be
d(\Delta) \sim \exp(c_S \sqrt{\Delta}) \qquad 1 \ll \Delta \ll N\,,
\ee
for some parameter $c_S$. 
One of the advantages of computing $\tilde{\alpha}$ instead of $\alpha(n,\ell)$ is that there is only one value to be computed for every $j$, while if we compute the growth parameter using \eqref{alphacriterion2} we are required to compute the value of $\alpha(n,\ell)$ for all polar states (there are roughly $\sim t^2/12$ such states). 

The $\cN = 2$ super-Virasoro minimal models all have elliptic genera of the form \eqref{thetafunctionproduct} (see App. \ref{app:minimalmodels} for the explicit expressions). Since for elliptic genera we have
\eq{Z_{\text{EG}}(\cC_1\otimes \cC _2)=Z_{\text{EG}}(\cC _1)Z_{\text{EG}}(\cC _2)\,,
}
tensor products of minimal models also have elliptic genera of the form \eqref{thetafunctionproduct}. Therefore, we are able to use the more convenient method and compute $\tilde{\alpha}$ to find the growth of the coefficients of the NS sector elliptic genus of $\operatorname{Sym}^N(\cC)$.

\section{Type IV theories}\label{app:unicorns}

In this section we give an explicit example of a weak Jacobi form that would be considered type IV (subject to an assumption that we specify later). The purpose is to illustrate that, if we want to prove that type IV theories never exist, we need to use a tool more than simply modularity.

The main tool we will use is that there exist ``spurious" slow-growing weak Jacobi forms at small $t, b$ that do not obey $t/b \in \mathbb{Z}$ ($t/b$ being an integer is a necessary condition for a weak Jacobi form to be the elliptic genus of a CFT with central charge $c_0=\frac{6b^2}{t}$ \cite{Belin:2020nmp}). We will then unwrap one of these (see (\ref{eq:zegwjf})), and take a linear combination of it with physical CFT elliptic genera in such a way that the marginal operators get cancelled. 

To give an explicit example, let us consider $t=19, b=3$. There is precisely a one-dimensional space of weak Jacobi forms with $t=19, b=3$ which obey criterion 2 \cite{Belin:2019rba, Keller:2020rwi}. Let us denote it as $\varphi_{19,3}(\tau, z)$. It has the following $q$-expansion:
\begin{align}
\begin{split}
    \varphi_{19,3}(\tau,z) &= (y^{-3} + 2 y^{-2} + 2 y^{-1} + 2 + 2y + 2y^2 + y^3)  \\&+ (-y^{-9} - 2y^{-8}-2y^{-7}-2y^{-6} -3y^{-5} -2y^{-4} +y^{-3} + 4y^{-2} + 5y^{-1} +4  \\&~~~~~~+5y+4y^2+y^3-2y^4-3y^5-2y^6-2y^7-2y^8-y^9)q + \mathcal{O}(q^2)\,.
    \end{split}
\end{align}
Since $t/b = 19/3$ is not an integer, this cannot be the elliptic genus of a physical CFT. However, if we unwrap it by a multiple of $3$, it can have the same $t, b$ as CFT elliptic genera with central charge $\frac{54}{19}$. There are two $\mathcal{N}=2$ SCFTs with this central charge -- the $A$- and $D$-series $k=36$ minimal models.\footnote{Note that even after unwrapping, $\varphi_{19,3}$ cannot literally be the elliptic genus of either of these two CFTs, because all of its charges will have a nontrivial greatest common divisor.}

From (3.8) of \cite{Belin:2020nmp}, we see the $A$- and $D$-series minimal models at $k=36$ have $t=684$, $b=18$ and $t=171$, $b=9$ respectively, and correspond to the following unwrapped weak Jacobi forms:
\begin{align}
\begin{split}
    Z_{A, k=36}(\tau, z) &= \frac{\theta_1(\tau,37z)}{\theta_1(\tau,z)}\,, \\
    Z_{D, k=36}(\tau, z) &= \frac{\theta_1(\tau, 18z)\theta_1(\tau,10z)}{\theta_1(\tau,9z)\theta_1(\tau,z)}\,.
    \end{split}
\end{align}
We will now unwrap $\varphi_{19,3}(\tau,z)$ by 6  and $Z_{D,k=36}$ by $2$ to make all three Jacobi forms have $t=684, b=18$ and take the following linear combination:
\begin{align}
\begin{split}
    &\varphi_{\text{Type IV}}(\tau, z) \coloneqq \varphi_{19,3}(\tau,6z) + Z_{A, k=36}(\tau, z) - Z_{D, k=36}(\tau,2z)  \\
    &\hspace{30pt}= y^{-18} + y^{-17} + y^{-15} + y^{-13} + 2y^{-12} + y^{-11} + y^{-9} + y^{-7} + 2y^{-6} + y^{-5}  \\&\hspace{30pt}~~~+ y^{-3} + y^{-1} + 1 + y + y^3 + y^5 + 2 y^6 + y^7 + y^9 + y^{11} + 2y^{12} + y^{13}  \\&\hspace{30pt}~~~ + y^{15} + y^{17} + y^{18} + \mathcal{O}(q)\,.
    \label{eq:typeivevilguy}
    \end{split}
\end{align}
Let us now argue that (\ref{eq:typeivevilguy}) -- if it corresponded to a bona fide CFT -- would be a type IV theory. Since it was written as a linear combination of weak Jacobi forms obeying criterion 2, (\ref{eq:typeivevilguy}) also obeys criterion 2.

To show that it has no marginal operators, we need full access to the half-BPS spectrum. In particular we need to show that (in the RR sector) there are no operators with charges $q^0 y^{(1 - \frac{L c_0}6) \frac tb}\bar{q}^0 \bar{y}^{(1 - \frac{L c_0}6) \frac tb}$ for $L=2, 3, \cdots$ (as well as any charge conjugations). For our example, $c_0=\frac{54}{19}, t=684, b=18$, this means we need to check charge 2 and 16 ($L>3$ is automatically vanishing by unitarity). 

Unfortunately, the elliptic genus only gives us a signed count. We have to make the assumption that if the $y^{\pm 2} q^0$ and $y^{\pm 16} q^0$ terms vanish in the elliptic genus, then the full unsigned half-BPS terms also vanish. If we make this assumption, we see that indeed, the desired terms vanish in (\ref{eq:typeivevilguy}).

Note that the function in (\ref{eq:typeivevilguy}) cannot correspond to a physical CFT because it would correspond to a SCFT with $c_0=\frac{54}{19}$, and $\mathcal{N}=2$ SCFTs with $c_0<3$ are fully classified by the minimal models. Since $\varphi_{\text{Type IV}}(\tau, z)$ is not a minimal model elliptic genus, it cannot be a physical CFT. Nonetheless, the example we gave here just illustrates that if we want to prove that type IV theories never exist, we need to use more than just modularity of the elliptic genus.

\bibliographystyle{ytphys}
\bibliography{ref}

\providecommand{\href}[2]{#2}\begingroup\raggedright\begin{thebibliography}{10}

\bibitem{Dijkgraaf:1996xw}
R.~Dijkgraaf, G.~W. Moore, E.~P. Verlinde, and H.~L. Verlinde, ``{Elliptic
  genera of symmetric products and second quantized strings},''
  \href{http://dx.doi.org/10.1007/s002200050087}{{\em Commun. Math. Phys.}
  {\bfseries 185} (1997) 197--209},
  \href{http://arxiv.org/abs/hep-th/9608096}{{\ttfamily arXiv:hep-th/9608096}}.

\bibitem{Pakman:2009zz}
A.~Pakman, L.~Rastelli, and S.~S. Razamat, ``{Diagrams for Symmetric Product
  Orbifolds},'' \href{http://dx.doi.org/10.1088/1126-6708/2009/10/034}{{\em
  JHEP} {\bfseries 10} (2009) 034},
\href{http://arxiv.org/abs/0905.3448}{{\ttfamily arXiv:0905.3448 [hep-th]}}.

\bibitem{Pakman:2009ab}
A.~Pakman, L.~Rastelli, and S.~S. Razamat, ``{Extremal Correlators and Hurwitz
  Numbers in Symmetric Product Orbifolds},''
  \href{http://dx.doi.org/10.1103/PhysRevD.80.086009}{{\em Phys. Rev. D}
  {\bfseries 80} (2009) 086009},
  \href{http://arxiv.org/abs/0905.3451}{{\ttfamily arXiv:0905.3451 [hep-th]}}.

\bibitem{Belin:2014fna}
A.~Belin, C.~A. Keller, and A.~Maloney, ``{String Universality for Permutation
  Orbifolds},'' \href{http://dx.doi.org/10.1103/PhysRevD.91.106005}{{\em Phys.
  Rev. D} {\bfseries 91} no.~10, (2015) 106005},
  \href{http://arxiv.org/abs/1412.7159}{{\ttfamily arXiv:1412.7159 [hep-th]}}.

\bibitem{Haehl:2014yla}
F.~M. Haehl and M.~Rangamani, ``{Permutation orbifolds and holography},''
  \href{http://dx.doi.org/10.1007/JHEP03(2015)163}{{\em JHEP} {\bfseries 03}
  (2015) 163},
\href{http://arxiv.org/abs/1412.2759}{{\ttfamily arXiv:1412.2759 [hep-th]}}.

\bibitem{Belin:2015hwa}
A.~Belin, C.~A. Keller, and A.~Maloney, ``{Permutation Orbifolds in the large N
  Limit},'' \href{http://dx.doi.org/10.1007/s00023-016-0529-y}{{\em Annales
  Henri Poincare} {\bfseries 18} (2017) 529--557},
\href{http://arxiv.org/abs/1509.01256}{{\ttfamily arXiv:1509.01256 [hep-th]}}.

\bibitem{Keller:2011xi}
C.~A. Keller, ``{Phase transitions in symmetric orbifold CFTs and
  universality},'' \href{http://dx.doi.org/10.1007/JHEP03(2011)114}{{\em JHEP}
  {\bfseries 03} (2011) 114},
\href{http://arxiv.org/abs/1101.4937}{{\ttfamily arXiv:1101.4937 [hep-th]}}.

\bibitem{Hartman:2014oaa}
T.~Hartman, C.~A. Keller, and B.~Stoica, ``{Universal Spectrum of 2d Conformal
  Field Theory in the Large c Limit},''
  \href{http://dx.doi.org/10.1007/JHEP09(2014)118}{{\em JHEP} {\bfseries 09}
  (2014) 118},
\href{http://arxiv.org/abs/1405.5137}{{\ttfamily arXiv:1405.5137 [hep-th]}}.

\bibitem{Gaberdiel:2015uca}
M.~R. Gaberdiel, C.~Peng, and I.~G. Zadeh, ``{Higgsing the stringy higher spin
  symmetry},'' \href{http://dx.doi.org/10.1007/JHEP10(2015)101}{{\em JHEP}
  {\bfseries 10} (2015) 101},
\href{http://arxiv.org/abs/1506.02045}{{\ttfamily arXiv:1506.02045 [hep-th]}}.

\bibitem{Apolo:2022fya}
L.~Apolo, A.~Belin, S.~Bintanja, A.~Castro, and C.~A. Keller, ``{Deforming
  Symmetric Product Orbifolds: A tale of moduli and higher spin currents},''
  \href{http://arxiv.org/abs/2204.07590}{{\ttfamily arXiv:2204.07590
  [hep-th]}}.

\bibitem{Maldacena:1997re}
J.~M. Maldacena, ``{The Large N limit of superconformal field theories and
  supergravity},'' \href{http://dx.doi.org/10.1023/A:1026654312961}{{\em
  Int.J.Theor.Phys.} {\bfseries 38} (1999) 1113--1133},
\href{http://arxiv.org/abs/hep-th/9711200}{{\ttfamily arXiv:hep-th/9711200
  [hep-th]}}.

\bibitem{Dijkgraaf:1998gf}
R.~Dijkgraaf, ``{Instanton strings and hyperKahler geometry},''
  \href{http://dx.doi.org/10.1016/S0550-3213(98)00869-4}{{\em Nucl. Phys. B}
  {\bfseries 543} (1999) 545--571},
  \href{http://arxiv.org/abs/hep-th/9810210}{{\ttfamily arXiv:hep-th/9810210}}.

\bibitem{Giveon:1998ns}
A.~Giveon, D.~Kutasov, and N.~Seiberg, ``{Comments on string theory on
  AdS(3)},'' \href{http://dx.doi.org/10.4310/ATMP.1998.v2.n4.a3}{{\em Adv.
  Theor. Math. Phys.} {\bfseries 2} (1998) 733--782},
  \href{http://arxiv.org/abs/hep-th/9806194}{{\ttfamily arXiv:hep-th/9806194}}.

\bibitem{Seiberg:1999xz}
N.~Seiberg and E.~Witten, ``{The D1 / D5 system and singular CFT},''
  \href{http://dx.doi.org/10.1088/1126-6708/1999/04/017}{{\em JHEP} {\bfseries
  04} (1999) 017},
\href{http://arxiv.org/abs/hep-th/9903224}{{\ttfamily arXiv:hep-th/9903224
  [hep-th]}}.

\bibitem{Argurio:2000tb}
R.~Argurio, A.~Giveon, and A.~Shomer, ``{Superstrings on AdS(3) and symmetric
  products},'' \href{http://dx.doi.org/10.1088/1126-6708/2000/12/003}{{\em
  JHEP} {\bfseries 12} (2000) 003},
  \href{http://arxiv.org/abs/hep-th/0009242}{{\ttfamily arXiv:hep-th/0009242}}.

\bibitem{David:2002wn}
J.~R. David, G.~Mandal, and S.~R. Wadia, ``{Microscopic formulation of black
  holes in string theory},''
  \href{http://dx.doi.org/10.1016/S0370-1573(02)00271-5}{{\em Phys. Rept.}
  {\bfseries 369} (2002) 549--686},
\href{http://arxiv.org/abs/hep-th/0203048}{{\ttfamily arXiv:hep-th/0203048
  [hep-th]}}.

\bibitem{Eberhardt:2018ouy}
L.~Eberhardt, M.~R. Gaberdiel, and R.~Gopakumar, ``{The Worldsheet Dual of the
  Symmetric Product CFT},''
  \href{http://dx.doi.org/10.1007/JHEP04(2019)103}{{\em JHEP} {\bfseries 04}
  (2019) 103},
\href{http://arxiv.org/abs/1812.01007}{{\ttfamily arXiv:1812.01007 [hep-th]}}.

\bibitem{Giribet:2018ada}
G.~Giribet, C.~Hull, M.~Kleban, M.~Porrati, and E.~Rabinovici, ``{Superstrings
  on AdS$_{3}$ at $k =$ 1},''
  \href{http://dx.doi.org/10.1007/JHEP08(2018)204}{{\em JHEP} {\bfseries 08}
  (2018) 204}, \href{http://arxiv.org/abs/1803.04420}{{\ttfamily
  arXiv:1803.04420 [hep-th]}}.

\bibitem{Eberhardt:2019ywk}
L.~Eberhardt, M.~R. Gaberdiel, and R.~Gopakumar, ``{Deriving the
  AdS$_{3}$/CFT$_{2}$ correspondence},''
  \href{http://dx.doi.org/10.1007/JHEP02(2020)136}{{\em JHEP} {\bfseries 02}
  (2020) 136}, \href{http://arxiv.org/abs/1911.00378}{{\ttfamily
  arXiv:1911.00378 [hep-th]}}.

\bibitem{Belin:2020nmp}
A.~Belin, N.~Benjamin, A.~Castro, S.~M. Harrison, and C.~A. Keller,
  ``{$\mathcal{N}=2$ Minimal Models: A Holographic Needle in a Symmetric
  Orbifold Haystack},''
  \href{http://dx.doi.org/10.21468/SciPostPhys.8.6.084}{{\em SciPost Phys.}
  {\bfseries 8} no.~6, (2020) 084},
  \href{http://arxiv.org/abs/2002.07819}{{\ttfamily arXiv:2002.07819
  [hep-th]}}.

\bibitem{Avery:2010er}
S.~G. Avery, B.~D. Chowdhury, and S.~D. Mathur, ``{Deforming the D1D5 CFT away
  from the orbifold point},''
  \href{http://dx.doi.org/10.1007/JHEP06(2010)031}{{\em JHEP} {\bfseries 06}
  (2010) 031},
\href{http://arxiv.org/abs/1002.3132}{{\ttfamily arXiv:1002.3132 [hep-th]}}.

\bibitem{Keller:2019yrr}
C.~A. Keller and I.~G. Zadeh, ``{Conformal Perturbation Theory for Twisted
  Fields},'' \href{http://dx.doi.org/10.1088/1751-8121/ab6b91}{{\em J. Phys. A}
  {\bfseries 53} no.~9, (2020) 095401},
  \href{http://arxiv.org/abs/1907.08207}{{\ttfamily arXiv:1907.08207
  [hep-th]}}.

\bibitem{Belin:2016yll}
A.~Belin, J.~de~Boer, J.~Kruthoff, B.~Michel, E.~Shaghoulian, and M.~Shyani,
  ``{Universality of sparse $d > 2$ conformal field theory at large $N$},''
  \href{http://dx.doi.org/10.1007/JHEP03(2017)067}{{\em JHEP} {\bfseries 03}
  (2017) 067},
\href{http://arxiv.org/abs/1610.06186}{{\ttfamily arXiv:1610.06186 [hep-th]}}.

\bibitem{Belin:2016dcu}
A.~Belin, B.~Freivogel, R.~Jefferson, and L.~Kabir, ``{Sub-AdS scale locality
  in AdS$_{3}$/CFT$_{2}$},''
  \href{http://dx.doi.org/10.1007/JHEP04(2017)147}{{\em JHEP} {\bfseries 04}
  (2017) 147}, \href{http://arxiv.org/abs/1611.08601}{{\ttfamily
  arXiv:1611.08601 [hep-th]}}.

\bibitem{Benjamin:2015hsa}
N.~Benjamin, M.~C.~N. Cheng, S.~Kachru, G.~W. Moore, and N.~M. Paquette,
  ``{Elliptic Genera and 3d Gravity},''
  \href{http://dx.doi.org/10.1007/s00023-016-0469-6}{{\em Annales Henri
  Poincare} {\bfseries 17} no.~10, (2016) 2623--2662},
\href{http://arxiv.org/abs/1503.04800}{{\ttfamily arXiv:1503.04800 [hep-th]}}.

\bibitem{Benjamin:2015vkc}
N.~Benjamin, S.~Kachru, C.~A. Keller, and N.~M. Paquette, ``{Emergent
  space-time and the supersymmetric index},''
  \href{http://dx.doi.org/10.1007/JHEP05(2016)158}{{\em JHEP} {\bfseries 05}
  (2016) 158},
\href{http://arxiv.org/abs/1512.00010}{{\ttfamily arXiv:1512.00010 [hep-th]}}.

\bibitem{Belin:2019rba}
A.~Belin, A.~Castro, C.~A. Keller, and B.~Muhlmann, ``{The Holographic
  Landscape of Symmetric Product Orbifolds},''
  \href{http://dx.doi.org/10.1007/JHEP01(2020)111}{{\em JHEP} {\bfseries 01}
  (2020) 111},
\href{http://arxiv.org/abs/1910.05342}{{\ttfamily arXiv:1910.05342 [hep-th]}}.

\bibitem{Belin:2019jqz}
A.~Belin, A.~Castro, C.~A. Keller, and B.~J. M{\"u}hlmann, ``{Siegel
  Paramodular Forms from Exponential Lifts: Slow versus Fast Growth},''
\href{http://arxiv.org/abs/1910.05353}{{\ttfamily arXiv:1910.05353 [hep-th]}}.

\bibitem{Keller:2020rwi}
C.~A. Keller and J.~M. Quinones, ``{On the space of slow growing weak Jacobi
  forms},'' \href{http://dx.doi.org/10.1016/j.jnt.2022.02.010}{{\em J. Number
  Theory} (2022) 730--750}, \href{http://arxiv.org/abs/2011.02611}{{\ttfamily
  arXiv:2011.02611 [math.NT]}}.

\bibitem{Maldacena:1999bp}
J.~M. Maldacena, G.~W. Moore, and A.~Strominger, ``{Counting BPS black holes in
  toroidal Type II string theory},''
  \href{http://arxiv.org/abs/hep-th/9903163}{{\ttfamily arXiv:hep-th/9903163}}.

\bibitem{Boucher:1986bh}
W.~Boucher, D.~Friedan, and A.~Kent, ``{Determinant Formulae and Unitarity for
  the N=2 Superconformal Algebras in Two-Dimensions or Exact Results on String
  Compactification},''
  \href{http://dx.doi.org/10.1016/0370-2693(86)90260-1}{{\em Phys. Lett. B}
  {\bfseries 172} (1986) 316}.

\bibitem{DiVecchia:1986fwg}
P.~Di~Vecchia, J.~L. Petersen, M.~Yu, and H.~B. Zheng, ``{Explicit Construction
  of Unitary Representations of the N=2 Superconformal Algebra},''
  \href{http://dx.doi.org/10.1016/0370-2693(86)91099-3}{{\em Phys. Lett. B}
  {\bfseries 174} (1986) 280--284}.

\bibitem{Gepner:1987qi}
D.~Gepner, ``{Space-Time Supersymmetry in Compactified String Theory and
  Superconformal Models},''
  \href{http://dx.doi.org/10.1016/0550-3213(88)90397-5}{{\em Nucl. Phys. B}
  {\bfseries 296} (1988) 757}.

\bibitem{Cecotti:1992rm}
S.~Cecotti and C.~Vafa, ``{On classification of N=2 supersymmetric theories},''
  \href{http://dx.doi.org/10.1007/BF02096804}{{\em Commun. Math. Phys.}
  {\bfseries 158} (1993) 569--644},
  \href{http://arxiv.org/abs/hep-th/9211097}{{\ttfamily arXiv:hep-th/9211097}}.

\bibitem{Gray:2008je}
O.~Gray, ``{On the complete classification of the unitary N=2 minimal
  superconformal field theories},''
  \href{http://dx.doi.org/10.1007/s00220-012-1478-z}{{\em Commun. Math. Phys.}
  {\bfseries 312} (2012) 611--654},
  \href{http://arxiv.org/abs/0812.1318}{{\ttfamily arXiv:0812.1318 [hep-th]}}.

\bibitem{Keller:2019suk}
C.~A. Keller and I.~G. Zadeh, ``{Lifting $\frac{1}{4}$-BPS States on K$3$ and
  Mathieu Moonshine},''
  \href{http://dx.doi.org/10.1007/s00220-020-03721-4}{{\em Commun. Math. Phys.}
  {\bfseries 377} no.~1, (2020) 225--257},
  \href{http://arxiv.org/abs/1905.00035}{{\ttfamily arXiv:1905.00035
  [hep-th]}}.

\bibitem{Guo:2020gxm}
B.~Guo and S.~D. Mathur, ``{Lifting at higher levels in the D1D5 CFT},''
  \href{http://dx.doi.org/10.1007/JHEP11(2020)145}{{\em JHEP} {\bfseries 11}
  (2020) 145}, \href{http://arxiv.org/abs/2008.01274}{{\ttfamily
  arXiv:2008.01274 [hep-th]}}.

\bibitem{Benjamin:2021zkn}
N.~Benjamin, C.~A. Keller, and I.~G. Zadeh, ``{Lifting 1/4-BPS states in
  AdS$_{3}$\texttimes{} S$^{3}$\texttimes{} T$^{4}$},''
  \href{http://dx.doi.org/10.1007/JHEP10(2021)089}{{\em JHEP} {\bfseries 10}
  (2021) 089}, \href{http://arxiv.org/abs/2107.00655}{{\ttfamily
  arXiv:2107.00655 [hep-th]}}.

\bibitem{Guo:2022ifr}
B.~Guo, M.~R.~R. Hughes, S.~D. Mathur, and M.~Mehta, ``{Universal lifting in
  the D1-D5 CFT},'' \href{http://arxiv.org/abs/2208.07409}{{\ttfamily
  arXiv:2208.07409 [hep-th]}}.

\bibitem{Benjamin:2016pil}
N.~Benjamin, ``{A Refined Count of BPS States in the D1/D5 System},''
  \href{http://dx.doi.org/10.1007/JHEP06(2017)028}{{\em JHEP} {\bfseries 06}
  (2017) 028}, \href{http://arxiv.org/abs/1610.07607}{{\ttfamily
  arXiv:1610.07607 [hep-th]}}.

\bibitem{Cappelli:1987xt}
A.~Cappelli, C.~Itzykson, and J.~B. Zuber, ``{The ADE Classification of Minimal
  and A1(1) Conformal Invariant Theories},''
\href{http://dx.doi.org/10.1007/BF01221394}{{\em Commun. Math. Phys.}
  {\bfseries 113} (1987) 1}.

\bibitem{deBoer:2008ss}
J.~de~Boer, J.~Manschot, K.~Papadodimas, and E.~Verlinde, ``{The Chiral ring of
  AdS(3)/CFT(2) and the attractor mechanism},''
  \href{http://dx.doi.org/10.1088/1126-6708/2009/03/030}{{\em JHEP} {\bfseries
  03} (2009) 030},
\href{http://arxiv.org/abs/0809.0507}{{\ttfamily arXiv:0809.0507 [hep-th]}}.

\bibitem{Dixon:1986qv}
L.~J. Dixon, D.~Friedan, E.~J. Martinec, and S.~H. Shenker, ``{The Conformal
  Field Theory of Orbifolds},''
  \href{http://dx.doi.org/10.1016/0550-3213(87)90676-6}{{\em Nucl. Phys. B}
  {\bfseries 282} (1987) 13--73}.

\bibitem{Lunin:2000yv}
O.~Lunin and S.~D. Mathur, ``{Correlation functions for $M^N / S(N)$
  orbifolds},'' \href{http://dx.doi.org/10.1007/s002200100431}{{\em Commun.
  Math. Phys.} {\bfseries 219} (2001) 399--442},
\href{http://arxiv.org/abs/hep-th/0006196}{{\ttfamily arXiv:hep-th/0006196
  [hep-th]}}.

\bibitem{Lunin:2001pw}
O.~Lunin and S.~D. Mathur, ``{Three point functions for $M(N) / S(N)$ orbifolds
  with N=4 supersymmetry},''
  \href{http://dx.doi.org/10.1007/s002200200638}{{\em Commun. Math. Phys.}
  {\bfseries 227} (2002) 385--419},
\href{http://arxiv.org/abs/hep-th/0103169}{{\ttfamily arXiv:hep-th/0103169
  [hep-th]}}.

\bibitem{Calabrese:2009qy}
P.~Calabrese and J.~Cardy, ``{Entanglement entropy and conformal field
  theory},'' \href{http://dx.doi.org/10.1088/1751-8113/42/50/504005}{{\em J.
  Phys. A} {\bfseries 42} (2009) 504005},
  \href{http://arxiv.org/abs/0905.4013}{{\ttfamily arXiv:0905.4013
  [cond-mat.stat-mech]}}.

\bibitem{Guo:2019ady}
B.~Guo and S.~D. Mathur, ``{Lifting of level-1 states in the D1D5 CFT},''
  \href{http://dx.doi.org/10.1007/JHEP03(2020)028}{{\em JHEP} {\bfseries 03}
  (2020) 028}, \href{http://arxiv.org/abs/1912.05567}{{\ttfamily
  arXiv:1912.05567 [hep-th]}}.

\bibitem{Gaberdiel:2010pz}
M.~R. Gaberdiel and R.~Gopakumar, ``{An AdS$_{3}$ Dual for Minimal Model
  CFTs},'' \href{http://dx.doi.org/10.1103/PhysRevD.83.066007}{{\em Phys. Rev.
  D} {\bfseries 83} (2011) 066007},
  \href{http://arxiv.org/abs/1011.2986}{{\ttfamily arXiv:1011.2986 [hep-th]}}.

\bibitem{Candu:2012jq}
C.~Candu and M.~R. Gaberdiel, ``{Supersymmetric holography on $AdS_3$},''
  \href{http://dx.doi.org/10.1007/JHEP09(2013)071}{{\em JHEP} {\bfseries 09}
  (2013) 071}, \href{http://arxiv.org/abs/1203.1939}{{\ttfamily arXiv:1203.1939
  [hep-th]}}.

\bibitem{Candu:2012tr}
C.~Candu and M.~R. Gaberdiel, ``{Duality in N=2 Minimal Model Holography},''
  \href{http://dx.doi.org/10.1007/JHEP02(2013)070}{{\em JHEP} {\bfseries 02}
  (2013) 070}, \href{http://arxiv.org/abs/1207.6646}{{\ttfamily arXiv:1207.6646
  [hep-th]}}.

\bibitem{Gaberdiel:2016xwo}
M.~R. Gaberdiel and M.~Kelm, ``{The symmetric orbifold of $ \mathcal{N}=2 $
  minimal models},'' \href{http://dx.doi.org/10.1007/JHEP07(2016)113}{{\em
  JHEP} {\bfseries 07} (2016) 113},
  \href{http://arxiv.org/abs/1604.03964}{{\ttfamily arXiv:1604.03964
  [hep-th]}}.

\bibitem{Lerche:1989uy}
W.~Lerche, C.~Vafa, and N.~P. Warner, ``{Chiral Rings in N=2 Superconformal
  Theories},'' \href{http://dx.doi.org/10.1016/0550-3213(89)90474-4}{{\em Nucl.
  Phys. B} {\bfseries 324} (1989) 427--474}.

\bibitem{Blumenhagen:2009zz}
R.~Blumenhagen and E.~Plauschinn, ``{Introduction to conformal field theory},''
\href{http://dx.doi.org/10.1007/978-3-642-00450-6}{{\em Lect.Notes Phys.}
  {\bfseries 779} (2009) 1--256}.

\bibitem{Schwimmer:1986mf}
A.~Schwimmer and N.~Seiberg, ``{Comments on the N=2, N=3, N=4 Superconformal
  Algebras in Two-Dimensions},''
  \href{http://dx.doi.org/10.1016/0370-2693(87)90566-1}{{\em Phys. Lett. B}
  {\bfseries 184} (1987) 191--196}.

\bibitem{Witten:1993jg}
E.~Witten, ``{On the Landau-Ginzburg description of N=2 minimal models},''
  \href{http://dx.doi.org/10.1142/S0217751X9400193X}{{\em Int. J. Mod. Phys. A}
  {\bfseries 9} (1994) 4783--4800},
  \href{http://arxiv.org/abs/hep-th/9304026}{{\ttfamily arXiv:hep-th/9304026}}.

\bibitem{DiFrancesco:1993dg}
P.~Di~Francesco and S.~Yankielowicz, ``{Ramond sector characters and N=2
  Landau-Ginzburg models},''
  \href{http://dx.doi.org/10.1016/0550-3213(93)90452-U}{{\em Nucl. Phys. B}
  {\bfseries 409} (1993) 186--210},
  \href{http://arxiv.org/abs/hep-th/9305037}{{\ttfamily arXiv:hep-th/9305037}}.

\bibitem{deBoer:1998us}
J.~de~Boer, ``{Large N elliptic genus and AdS / CFT correspondence},''
  \href{http://dx.doi.org/10.1088/1126-6708/1999/05/017}{{\em JHEP} {\bfseries
  05} (1999) 017},
\href{http://arxiv.org/abs/hep-th/9812240}{{\ttfamily arXiv:hep-th/9812240
  [hep-th]}}.

\end{thebibliography}\endgroup

\end{document}